\documentclass[aps,prd,preprintnumbers,showpacs,showkeys,nofootinbib,superscriptaddress,fleqn,floatfix,tightenlines,10pt]{revtex4-2} 
\usepackage{amsmath,amsfonts,amssymb,amscd,amsxtra,amsthm}
\usepackage{dsfont}
\usepackage{graphicx}  
\usepackage{epstopdf}
\usepackage{dcolumn}  
\usepackage{bm}          
\usepackage{slashed}
\usepackage[utf8]{inputenc}
\usepackage{hyperref}
\usepackage{booktabs}
\usepackage[normalem]{ulem} 
\usepackage[dvipsnames]{xcolor} 
\usepackage{enumitem}
\usepackage{array}
\usepackage{slashed}
\usepackage{tikz}
\usepackage{float}
\usepackage{multirow}
\usepackage{booktabs} 
\usepackage{slashed}

\renewcommand\sout{\bgroup \color{red} \ULdepth=-.5ex \ULset}

\begin{document}
\preprint{INHA-NTG-04/2025}
\title{\Large Double-strangeness hidden-charm pentaquarks}
\author{Samson Clymton}
\email[E-mail: ]{samson.clymton@apctp.org}
\affiliation{Asia Pacific Center for Theoretical Physics (APCTP),
  Pohang, Gyeongbuk 37673, Republic of Korea} 

\author{Hyun-Chul Kim}
\email[E-mail: ]{hchkim@inha.ac.kr}
\affiliation{Department of Physics, Inha University,
  Incheon 22212, Republic of Korea}
\affiliation{Physics Research Institute, Inha University, Incheon
  22212, Republic of Korea} 
\affiliation{School of Physics, Korea Institute for Advanced Study 
  (KIAS), Seoul 02455, Republic of Korea}

\author{Terry Mart}
\email[E-mail: ]{terry.mart@sci.ui.ac.id}
\affiliation{Departemen Fisika, FMIPA, Universitas Indonesia, Depok
  16424, Indonesia} 
\date{\today}
\begin{abstract}
We investigate the possible existence of double-strangeness
hidden-charm pentaquark states, denoted as $P_{c\bar{c}ss}$, within an
off-shell coupled-channel formalism. Eleven meson--baryon channels
with total strangeness $S = -2$ are constructed by combining charmed
mesons and singly charmed baryons. The two-body scattering amplitudes
are derived from an effective Lagrangian that respects heavy-quark
spin symmetry, hidden local symmetry, and flavor SU(3) symmetry. The
Bethe--Salpeter equation is solved using the Blankenbecler--Sugar
reduction scheme, and resonances are identified as poles in the
scattering amplitudes on the complex energy plane. 
We find five negative-parity $P_{c\bar{c}ss}$ states with spins $J =
1/2$, $3/2$, and $5/2$, all located below their relevant
thresholds. Three positive-parity states are also found: two with $J =
1/2$ and one with $J = 3/2$, lying above the thresholds with
substantial widths. The coupling strengths of each resonance to
relevant meson--baryon channels are extracted. The sensitivity of the
results to the cutoff parameter $\Lambda_0 = \Lambda - m$ is
examined. These results provide theoretical
predictions that may assist future experimental searches for
$P_{c\bar{c}ss}$ states in the $J/\psi\,\Xi$ channel.
\end{abstract} 
\maketitle
\section{Introduction}
The observation of hidden-charm pentaquark states~\cite{LHCb:2015yax,
  LHCb:2019kea, LHCb:2021chn} has generated significant experimental
and theoretical interest in heavy pentaquark states (see the recent
reviews~\cite{Esposito:2016noz, Chen:2016spr, Meng:2022ozq,
  Chen:2022asf} and references therein). Subsequently, the LHCb
Collaboration reported the existence of a neutral hidden-charm
pentaquark with strangeness, denoted as
$P_{c\bar{c}s}(4459)$~\cite{LHCb:2021chn, LHCb:2022ogu}. More
recently, the Belle Collaboration confirmed the existence of
$P_{c\bar{c}s}(4459)$ but reported a slightly larger mass,
$M_{P_{c\bar{c}s}} = (4471.7 \pm 4.8 \pm 0.6)$~MeV/$c^2$, and a decay
width of $\Gamma = (21.9 \pm 13.1 \pm 2.7)$~MeV~\cite{Belle:2025pey}.  
Given that the CMS and LHCb Collaborations have recently reported the
decays $\Lambda_b^0 \to J/\psi \Xi^- K^+$ and $\Xi_b^0 \to J/\psi
\Xi^- \pi^+$~\cite{CMS:2024vnm, LHCb:2025lhk}, the possible existence
of hidden-charm pentaquark states with strangeness $S = -2$, denoted
as $P_{c\bar{c}ss}$, is anticipated. There have already been
theoretical predictions for such double-strangeness hidden-charm
pentaquarks~\cite{Wang:2020bjt, Ortega:2022uyu, Marse-Valera:2022khy,
  Roca:2024nsi, Marse-Valera:2024apc}. 

In our previous work, we studied the production of hidden-charm
pentaquark states with strangeness $S = 0$ and $S = -1$, using an
off-shell coupled-channel approach to two-body charmed hadronic
scattering~\cite{Clymton:2024fbf, Clymton:2025hez}. We first
constructed the two-body kernel Feynman amplitudes for the relevant
channels using an effective Lagrangian that respects heavy-quark
spin--flavor symmetry, hidden local symmetry, and chiral symmetry. We
then solved the coupled-channel scattering integral equations
involving seven different channels for the non-strange $P_{c\bar{c}}$
states and nine different channels for the $P_{c\bar{c}s}$ states. The
hidden-charm pentaquark states emerge as poles in the complex energy
plane. 

We found four hidden-charm pentaquark states with negative parity,
which we associated with the $P_{c\bar{c}}$ states observed by the
LHCb Collaboration: $P_{c\bar{c}}(4312)$, $P_{c\bar{c}}(4380)$,
$P_{c\bar{c}}(4440)$, and $P_{c\bar{c}}(4457)$. In addition, we
predicted the existence of two more negative-parity and two
positive-parity $P_{c\bar{c}}$ states. We also explained the absence
of a $P_{c\bar{c}}$ signal in the GlueX experiment for $J/\psi N$
photoproduction~\cite{GlueX:2019mkq}: destructive interference in
$J/\psi N$ scattering, combined with suppression due to dominant
positive-parity contributions, weakens the hidden-charm pentaquark
signals in the $J/\psi N$ channel~\cite{Clymton:2024fbf}.  
Regarding the hidden-charm pentaquarks with $S = -1$, we identified
seven $P_{c\bar{c}s}$ states with negative parity, among which
$P_{c\bar{c}s}(4459)$, located below the $\bar{D}^*\Xi_c$ threshold,
corresponds to the state observed by the LHCb
Collaboration~\cite{LHCb:2021chn, LHCb:2022ogu}. Interestingly, we
also found another state, $P_{c\bar{c}s}(4472)$, below the
$\bar{D}^*\Xi_c$ threshold, whose mass matches that reported by the
Belle Collaboration. Although this state has been considered identical
to $P_{c\bar{c}s}(4459)$, it may in fact represent a separate
resonance. 

\begin{figure}[htp]
  \centering
  \includegraphics[scale=0.35]{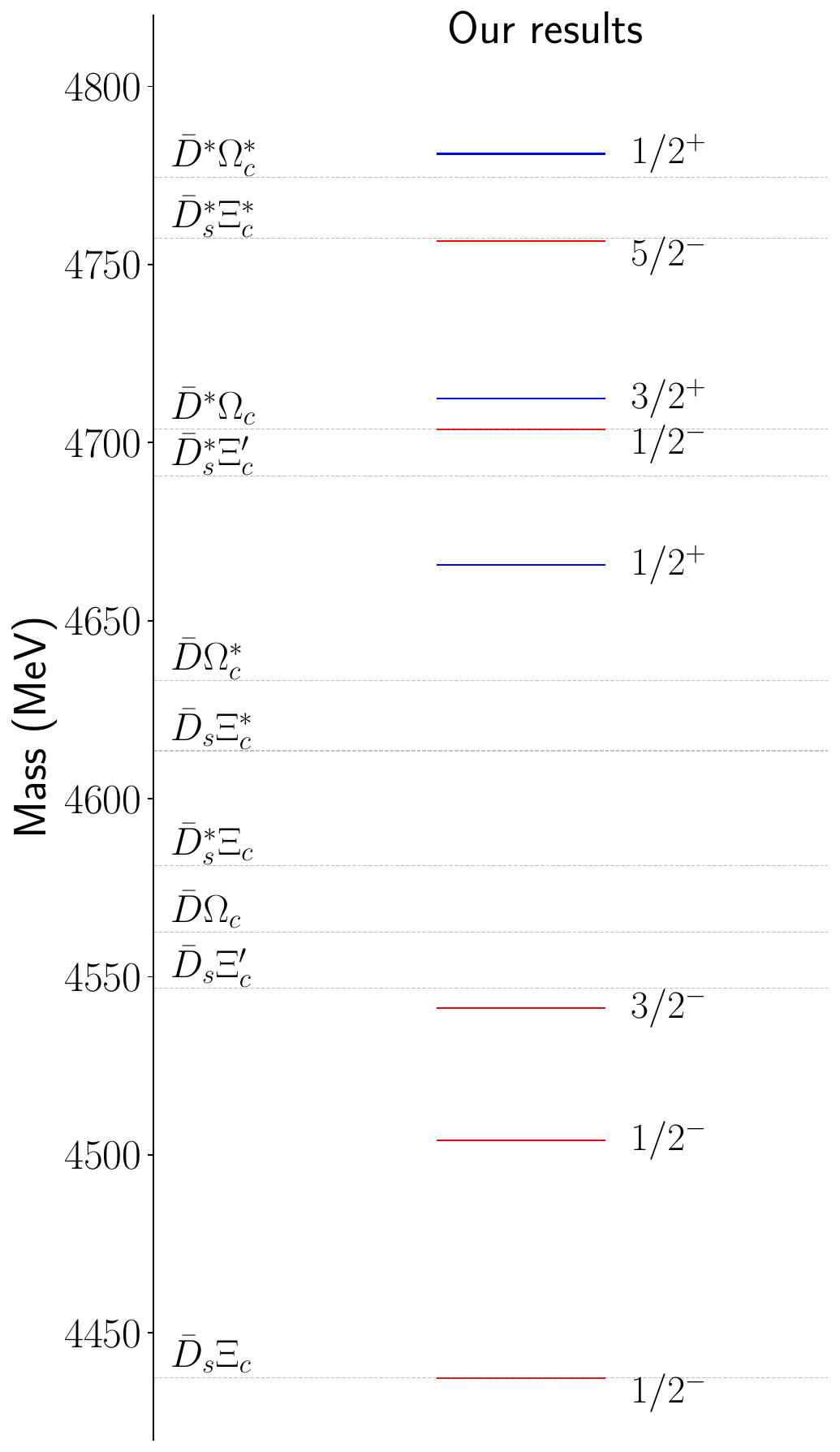}
  \caption{Predicted $P_{c\bar{c}ss}$ states from the present
    work. Red lines denote double-strangeness hidden-charm
    pentaquark states with negative parity, whereas blue lines
    represent those with positive parity.}  
  \label{fig:1}  
\end{figure}
In the present work, we extend the off-shell coupled-channel formalism
to study double-strangeness hidden-charm pentaquark
states. Figure~\ref{fig:1} summarizes the predictions for the
$P_{c\bar{c}ss}$ pentaquarks obtained in this study. We have found
five resonances with negative parity and three with positive parity,
covering spins $1/2$, $3/2$, and $5/2$. Notably, the negative-parity
states lie below the corresponding thresholds, while the
positive-parity ones are located above the thresholds. 

The organization of the present work is as follows: 
In Sec.~\ref{sec:2}, we describe the off-shell coupled-channel
formalism. We begin by showing how the two-body kernel Feynman amplitudes
are derived from an effective Lagrangian and proceed to solve the
scattering integral equations. 
In Sec.~\ref{sec:3}, we discuss the results for the
double-strangeness hidden-charm pentaquarks. 
The final section is devoted to the summary and conclusions.
\section{Coupled-channel formalism\label{sec:2}} 
\begin{figure}[htbp]
  \centering
  \includegraphics[scale=1.0]{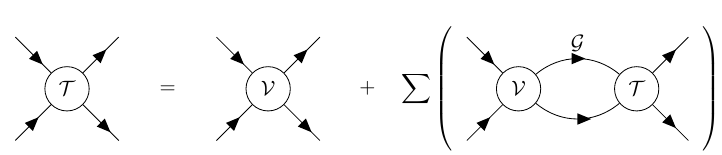}
  \caption{Graphical representation of the coupled 
          integral equation.}  
  \label{fig:2}
\end{figure}

The scattering amplitude can be expressed as 
\begin{align}
\mathcal{S}_{fi} = \delta_{fi} - i (2\pi)^4 \delta(P_f - P_i)
  \mathcal{T}_{fi}, 
\end{align}
where $P_i$ and $P_f$ denote the total four-momenta of the initial
and final states, respectively. The transition amplitudes
$\mathcal{T}_{fi}$ are obtained from the Bethe--Salpeter equation with
the two-body Feynman kernel amplitudes: 
\begin{align}
\mathcal{T}_{fi} (p',p;s) =\, \mathcal{V}_{fi}(p',p;s) 
+ \frac{1}{(2\pi)^4}\sum_k \int d^4q \,
\mathcal{V}_{fk}(p',q;s)\,\mathcal{G}_{k}(q;s)\,\mathcal{T}_{ki}(q,p;s),  
\label{eq:BSE}
\end{align}
where $p$ and $p'$ are the relative four-momenta of
the initial and final states, and $q$ represents the off-shell 
momentum for intermediate states in the center-of-mass (CM)
frame. The variable $s \equiv P_i^2 = P_f^2$ denotes the
square of the total energy and corresponds to one of the Mandelstam
variables. The indices $i$, $f$, and $k$ refer to the initial, final,
and intermediate states, respectively. Since we consider coupled
channels relevant to the production of $P_{c\bar{c}ss}$ states, the
index $k$ runs over all relevant intermediate channels. 
The coupled integral scattering equations given in Eq.~\eqref{eq:BSE} are 
illustrated schematically in Fig.~\ref{fig:2}. 

To simplify the challenges posed by the four-dimensional integral
equations, we adopt a three-dimensional reduction approach. Among the 
various methods available, we employ the Blankenbecler--Sugar
formalism~\cite{Blankenbecler:1965gx, Aaron:1968aoz}, which
characterizes the two-body propagator as  
\begin{align}
  \mathcal{G}_k(q) =\;
  \delta\left(q_0-\frac{E_{k1}(\bm{q})-E_{k2}(\bm{q})}{2}\right)
  \frac{\pi}{E_{k1}(\bm{q})E_{k2}(\bm{q})}
  \frac{E_k(\bm{q})}{s-E_k^2(\bm{q})}.
\label{eq:tbprop}
\end{align}
Here, $E_k$ denotes the total on-shell energy of the
intermediate state, defined as $E_k = E_{k1} + E_{k2}$, and $\bm{q}$
represents the three-momentum of the intermediate state. It is
important to note that the spinor contributions from the
meson--baryon propagator $G_k$ have been absorbed into the matrix
elements of $\mathcal{V}$ and $\mathcal{T}$. 
Using Eq.~\eqref{eq:tbprop}, we obtain the following coupled integral
equations:      
\begin{align}
  \mathcal{T}_{fi} (\bm{p}',\bm{p}) =\, \mathcal{V}_{fi}
  (\bm{p}',\bm{p}) 
  +\frac{1}{(2\pi)^3}\sum_k\int \frac{d^3q}{2E_{k1}(\bm{q})E_{k2}
  (\bm{q})} \mathcal{V}_{fk}(\bm{p}',\bm{q})\frac{E_k
  (\bm{q})}{s-E_k^2(\bm{q})+i\varepsilon} 
  \mathcal{T}_{ki}(\bm{q},\bm{p}),
  \label{eq:BS-3d}
\end{align}
where $\bm{p}$ and $\bm{p}'$ denote the relative three-momenta of the
initial and final states, respectively, in the CM frame. 

\begin{figure}[htbp]
  \centering
  \includegraphics[scale=0.35]{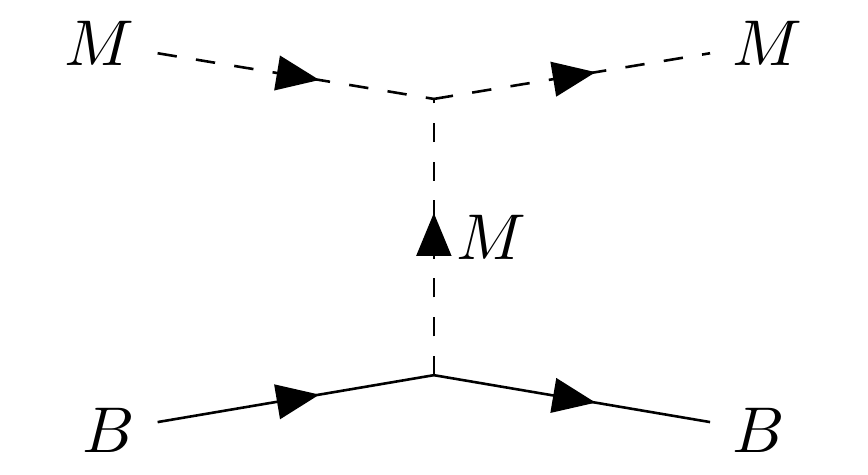}
  \caption{$t$-channel meson-exchange diagrams. $M$ and $B$ denote the
          meson and baryon, respectively.} 
  \label{fig:3}
\end{figure}
To investigate the pentaquark $P_{c\bar{c}ss}$'s, we construct two-body
coupled channels by combining the charmed meson triplet with the singly
charmed baryon antitriplet and sextet, while maintaining a total
strangeness number of $S=-2$. We also include the $J/\psi \Xi$
channel, considering the fact that both the $P_{c\bar{c}}$ and
$P_{c\bar{c}s}$ were observed in the invariant masses of $J/\psi N$
and $J/\psi \Lambda$, respectively. 
Thus, we include eleven distinct channels: $J/\psi\,\Xi$,
$\bar{D}_s\,\Xi_c$, $\bar{D}_s\,\Xi_c'$, $\bar{D}\,\Omega_c$,
$\bar{D}_s^*\,\Xi_c$, $\bar{D}_s\,\Xi_c^*$, $\bar{D}\,\Omega_c^*$,
$\bar{D}_s^*\,\Xi_c'$, 
$\bar{D}^*\,\Omega_c$, $\bar{D}_s^*\,\Xi_c^*$, and
$\bar{D}^*\,\Omega_c^*$. The two-body kernel Feynman amplitudes are
constructed from tree-level one-meson-exchange diagrams, as
illustrated in Fig.~\ref{fig:3}. 

We exclude pole diagrams in the $s$-channel, since we want to see how
the $P_{c\bar{c}ss}$'s can be dynamically generated.
The $u$-channel diagrams, which involve the exchange of
doubly-charmed baryons with masses of approximately 3.5~GeV, are 
significantly suppressed in magnitude compared to the $t$-channel
contributions and are thus also neglected in the present analysis.

We determine the vertex interactions using an effective Lagrangian
that satisfies heavy-quark spin symmetry, hidden local gauge
symmetry, and flavor SU(3) symmetry~\cite{Casalbuoni:1996pg}, which
are expressed as 
\begin{align}
  \mathcal{L}_{PP\mathbb{V}} &= -i\frac{\beta g_V}{\sqrt{2}}\,
  P^{\dagger}_a  \overleftrightarrow{\partial_\mu} P_b\,
   \mathbb{V}^\mu_{ba} +i\frac{\beta g_V}{\sqrt{2}}\,P'^{\dagger}_a
                               \overleftrightarrow{\partial_\mu}
                               P'_b\, \mathbb{V}^\mu_{ab},\\ 
    \mathcal{L}_{PP\sigma} &= -2g_\sigma M P^\dagger_a \sigma P_a
  -2g_\sigma M P'^\dagger_a\sigma P'_a ,\\     %
    \mathcal{L}_{P^*P^*\mathbb{P}} &= -\frac{g}{f_\pi}
  \epsilon^{\mu\nu\alpha\beta}P^{*\dagger}_{a\nu}\,
   \overleftrightarrow{\partial_\mu}\,
P^*_{b\beta}\partial_\alpha \mathbb{P}_{ba} -\frac{g}{f_\pi}
  \epsilon^{\mu\nu\alpha\beta}  P'^{*\dagger}_{a\nu}\,
 \overleftrightarrow{\partial_\mu}\,
 P'^*_{b\beta}\partial_\alpha
                                     \mathbb{P}_{ab} ,\\     %
    \mathcal{L}_{P^*P^*\mathbb{V}} & = i\frac{\beta g_V}{\sqrt{2}} \,
 P^{*\dagger}_{a\nu}  \overleftrightarrow{\partial_\mu}
 P^{*\nu}_b \mathbb{V}_{ba}^\mu + i2\sqrt{2} \lambda g_VM^*
 P^{*\dagger}_{a\mu}  P^*_{b\nu}\mathbb{V}_{ba}^{\mu\nu}\cr
  &\;\;\;\;-i\frac{\beta g_V}{\sqrt{2}} \,P'^{*\dagger}_{a\nu}
    \overleftrightarrow{\partial_\mu} P'^{*\nu}_b
    \mathbb{V}_{ab}^\mu-i2\sqrt{2}\lambda g_VM^*
    P'^{*\dagger}_{a\mu} P'^*_{b\nu}\mathbb{V}_{ab}^{\mu\nu} ,\\
  \mathcal{L}_{P^*P^*\sigma} &= 2g_\sigma M^* P^{*\dagger}_{a\mu}
\sigma P^{*\mu}_a+2g_\sigma M^*
 P'^{*\dagger}_{a\mu}\sigma P'^{*\mu}_a ,\\    %
    \mathcal{L}_{P^*P\mathbb{P}} &= -\frac{2g}{f_\pi}  \sqrt{MM^*}\,
      \left( P^{\dagger}_a P^*_{b\mu}+P^{*\dagger}_{a\mu} P_b\right)\,
 \partial^\mu \mathbb{P}_{ba}+\frac{2g}{f_\pi} \sqrt{MM^*}\,
    \left(P'^{\dagger}_a P'^*_{b\mu}+P'^{*\dagger}_{a\mu} P'_b\right)\,
  \partial^\mu \mathbb{P}_{ab},\\
  \mathcal{L}_{P^*P\mathbb{V}} &= -i\sqrt{2}\lambda g_V\,
\epsilon^{\beta\alpha\mu\nu} \left(P^{\dagger}_a
\overleftrightarrow{\partial_\beta} P^*_{b\alpha} +
P^{*\dagger}_{a\alpha}  \overleftrightarrow{\partial_\beta}
  P_b\right)\,\left(\partial_\mu\mathbb{V}_{\nu}\right)_{ba}\cr 
 &\;\;\;\;-i\sqrt{2}\lambda g_V\, \epsilon^{\beta\alpha\mu\nu}
   \left(P'^{\dagger}_a\overleftrightarrow{\partial_\beta}
   P'^*_{b\alpha}+P'^{*\dagger}_{a\alpha}
   \overleftrightarrow{\partial_\beta}P'_b\right)\,
   \left(\partial_\mu\mathbb{V}_{\nu}\right)_{ab},
\end{align}
where $\overleftrightarrow{\partial} =
\overrightarrow{\partial}-\overleftarrow{\partial}$. The  
$\sigma$ stands for the lowest-lying isoscalar-scalar meson,
$f_0$. Heavy mesons and anti-heavy mesons $P^{(*)}$ and $P'^{(*)}$ are
given as
\begin{align}
  P = \left(D^0,D^+,D_s^+\right), \hspace{0.5 cm}
  P^*_\mu =\left(D^{*0}_\mu,D^{*+}_\mu,D_{s\mu}^{*+}\right),
  \hspace{0.5 cm} P' =(\bar{D}^0,\,D^-,\,D_s^-),
  \hspace{0.5 cm} P'^*_\mu =(\bar{D}^{*0}_\mu,\,D^{*-}_\mu,\,D^{*-}_{s\mu}),
\end{align}
while the light pseudoscalar and vector mesons
are written as 
\begin{align}
    \mathbb{P} = 
    \begin{pmatrix}
        \frac{1}{\sqrt{2}} \pi^0+\frac{1}{\sqrt{6}}\eta & \pi^+ & K^+\\
        \pi^- & -\frac{1}{\sqrt{2}} \pi^0+\frac{1}{\sqrt{6}}\eta & K^0\\
        K^- & \bar{K}^0 & -\frac{2}{\sqrt{6}}\eta
    \end{pmatrix},\;\;\;\;
    \mathbb{V}_\mu = \begin{pmatrix}
        \frac{1}{\sqrt{2}} \rho^0_\mu+\frac{1}{\sqrt{2}}\omega_\mu &
        \rho_\mu^+ & K_\mu^{*+}\\
        \rho_\mu^- & -\frac{1}{\sqrt{2}} \rho_\mu^0+\frac{1}{\sqrt{2}}
        \omega_\mu & K_\mu^{*0} \\
        K_\mu^{*-} & \bar{K}^{*0}_\mu & \phi_\mu
    \end{pmatrix}.
\end{align}

We determine the coupling constants in the Lagrangian as
follows~\cite{Isola:2003fh}: $g = 0.59 \pm 0.07 \pm 0.01$, which is 
established through experimental measurements of the full width for
$D^{*+}$; $g_V = m_\rho / f_\pi \approx 5.8$, which is derived by
using the Kawarabayashi-Suzuki-Riazuddin-Fayyazuddin (KSRF) relation
with $f_\pi = 132~\mathrm{MeV}$; $\beta \approx
0.9$~\cite{Kawarabayashi:1966kd, Riazuddin:1966sw}, which stems from
the vector meson dominance in heavy meson radiative decay;
and $\lambda = -0.56~\mathrm{GeV}^{-1}$, which is extracted from
light-cone sum rules and lattice QCD. It is important to
note that the sign convention for $\lambda$ differs from that in 
Ref.~\cite{Isola:2003fh}, as we utilize the same phase for heavy
vector mesons found in Ref.~\cite{Casalbuoni:1996pg}.  
For the sigma meson, the coupling constant is utilized to assess the
$2\pi$ transition of $D_s(1^+)$ in Ref.~\cite{Bardeen:2003kt}. The
lowest isoscalar-scalar meson coupling is expressed as $g_\sigma =
g_\pi/2\sqrt{6}$ with $g_\pi = 3.73$. 

For the heavy baryon effective Lagrangian, we utilize the framework
from Ref.~\cite{Liu:2011xc}, which incorporates a more comprehensive
Lagrangian formulation as outlined in Ref.~\cite{Yan:1992gz}. The
baryonic interaction vertices within the tree-level meson-exchange
diagrams are characterized by the following effective Lagrangian: 
\begin{align}
    \mathcal{L}_{B_{\bar{3}}B_{\bar{3}}\mathbb{V}}& =
 \frac{i\beta_{\bar{3}}g_V}{2\sqrt{2}M_{\bar{3}}}
  \left(\bar{B}_{\bar{3}}\overleftrightarrow{\partial_\mu}
  \mathbb{V}^\mu B_{\bar{3}}\right) ,\\
  \mathcal{L}_{B_{\bar{3}}B_{\bar{3}}\sigma}&=l_{\bar{3}}
 \left(\bar{B}_{\bar{3}}\sigma B_{\bar{3}}\right) ,\\ 
    \mathcal{L}_{B_6B_6\mathbb{P}}&= i\frac{g_1}{2f_\pi M_6}\bar{B}_{6}\gamma_5\left(\gamma^\alpha\gamma^\beta-g^{\alpha\beta}\right)\overleftrightarrow{\partial_\alpha}\partial_\beta\mathbb{P} B_{6} ,\\
    \mathcal{L}_{B_6B_6\mathbb{V}}&= -i\frac{\beta_6 g_V}{2\sqrt{2}M_6}\left(\bar{B}_{6}\overleftrightarrow{\partial_\alpha}\mathbb{V}^\alpha B_{6}\right)-\frac{i\lambda_6g_V}{3\sqrt{2}}\left(\bar{B}_{6}\gamma_\mu\gamma_\nu \mathbb{V}^{\mu\nu}B_{6}\right) ,\\
    \mathcal{L}_{B_6B_6\sigma}&= -l_6\left(\bar{B}_{6}\sigma B_6\right) ,\\
    \mathcal{L}_{B_6^*B_6^*\mathbb{P}}&= \frac{3g_1}{4f_\pi M_6^*}\epsilon^{\mu\nu\alpha\beta}\left(\bar{B}_{6\mu}^*\overleftrightarrow{\partial_\nu}\partial_\alpha\mathbb{P} B_{6\beta}^*\right) ,\\
    \mathcal{L}_{B_6^*B_6^*\mathbb{V}}&= i\frac{\beta_6 g_V}{2\sqrt{2}M_6^*}\left(\bar{B}_{6\mu}^*\overleftrightarrow{\partial_\alpha}\mathbb{V}^\alpha B_{6}^{*\mu}\right)+\frac{i\lambda_6g_V}{\sqrt{2}} \left(\bar{B}^*_{6\mu} \mathbb{V}^{\mu\nu}B^*_{6\nu}\right) ,\\
    \mathcal{L}_{B_6^*B_6^*\sigma}&= l_6\left(\bar{B}^*_{6\mu}\sigma B^{*\mu}_6\right) ,\\
    \mathcal{L}_{B_6B_6^*\mathbb{P}}&= \frac{g_1}{4f_\pi}\sqrt{\frac{3}{M_6^*M_6}}\epsilon^{\mu\nu\alpha\beta}\left[\left(\bar{B}_{6}\gamma_5\gamma_\mu\overleftrightarrow{\partial_\nu} \partial_\alpha\mathbb{P} B_{6\beta}^*\right)+\left(\bar{B}_{6\mu}^*\gamma_5\gamma_\nu\overleftrightarrow{\partial_\alpha} \partial_\beta\mathbb{P}B_6 \right)\right] ,\\
    \mathcal{L}_{B_6B_6^*\mathbb{V}}&= \frac{i\lambda_6g_V}{\sqrt{6}}\left[\bar{B}_{6}\gamma_5\left(\gamma_\mu +\frac{i\overleftrightarrow{\partial_\mu}}{2\sqrt{M_6^*M_6}}\right) \mathbb{V}^{\mu\nu}B^*_{6\nu}+\bar{B}_{6\mu}^*\gamma_5\left(\gamma_\nu -\frac{i\overleftrightarrow{\partial_\nu}}{2\sqrt{M_6^*M_6}}\right) \mathbb{V}^{\mu\nu}B_{6}\right] ,\\
    \mathcal{L}_{B_6B_{\bar{3}}\mathbb{P}}&= -\frac{g_4}{\sqrt{3}f_\pi}\left[\bar{B}_6\gamma_5\left(\gamma_\mu +\frac{i\overleftrightarrow{\partial_\mu}}{2\sqrt{M_6M_{\bar{3}}}} \right) \partial^\mu\mathbb{P} B_{\bar{3}}+\bar{B}_{\bar{3}}\gamma_5\left(\gamma_\mu -\frac{i\overleftrightarrow{\partial_\mu}}{2\sqrt{M_6M_{\bar{3}}}} \right) \partial^\mu\mathbb{P}\,B_{6}\right] ,\\
    \mathcal{L}_{B_6B_{\bar{3}}\mathbb{V}}&= i\frac{\lambda_{6\bar{3}}\,g_V}{\sqrt{6M_6M_{\bar{3}}}}\epsilon^{\mu\nu\alpha\beta}\left[\left(\bar{B}_{6}\gamma_5\gamma_\mu \overleftrightarrow{\partial_\nu}\partial_\alpha\mathbb{V}_{\beta} B_{\bar{3}}\right)+\left(\bar{B}_{\bar{3}}\gamma_5\gamma_\mu \overleftrightarrow{\partial_\nu}\partial_\alpha\mathbb{V}_{\beta} B_{6}\right)\right] ,\\
    \mathcal{L}_{B_6^*B_{\bar{3}}\mathbb{P}}&= -\frac{g_4}{f_\pi}\left[\left(\bar{B}^*_{6\mu}\partial^\mu\mathbb{P} B_{\bar{3}}\right)+\left(\bar{B}_{\bar{3}}\partial^\mu\mathbb{P} B^*_{6\mu}\right)\right] ,\\
    \mathcal{L}_{B_6^*B_{\bar{3}}\mathbb{V}}&= i\frac{\lambda_{6\bar{3}}\,g_V}{\sqrt{2M_6^*M_{\bar{3}}}}\epsilon^{\mu\nu\alpha\beta}\left[\left(\bar{B}^*_{6\mu}\overleftrightarrow{\partial_\nu}\partial_\alpha\mathbb{V}_{\beta} B_{\bar{3}}\right)+\left(\bar{B}_{\bar{3}}\overleftrightarrow{\partial_\nu}\partial_\alpha\mathbb{V}_{\beta} B^*_{6\mu}\right)\right] ,
\end{align}
with heavy baryon fields expressed as
\begin{align}
    &B_{\bar{3}}=
    \begin{pmatrix}
        0 & \Lambda_c^+ & \Xi_c^+\\
        -\Lambda_c^+ & 0 & \Xi_c^0\\
        -\Xi_c^+ & -\Xi_c^0 & 0
    \end{pmatrix},\;\;
    B_6 =
    \begin{pmatrix}
        \Sigma_c^{++} & \frac{1}{\sqrt{2}}\Sigma_c^+ & \frac{1}{\sqrt{2}}\Xi{'}_c^+\\
        \frac{1}{\sqrt{2}}\Sigma_c^+ & \Sigma_c^0 & \frac{1}{\sqrt{2}}\Xi{'}_c^0\\
        \frac{1}{\sqrt{2}}\Xi{'}_c^+ & \frac{1}{\sqrt{2}}\Xi{'}_c^0 & \Omega_c^0
    \end{pmatrix},\;\;
    B_6^* =
    \begin{pmatrix}
        \Sigma_c^{*++} & \frac{1}{\sqrt{2}}\Sigma_c^{*+} & \frac{1}{\sqrt{2}}\Xi_c^{*+}\\
        \frac{1}{\sqrt{2}}\Sigma_c^{*+} & \Sigma_c^{*0} & \frac{1}{\sqrt{2}}\Xi_c^{*0}\\
        \frac{1}{\sqrt{2}}\Xi_c^{*+} & \frac{1}{\sqrt{2}}\Xi_c^{*0} & \Omega_c^{*0}
    \end{pmatrix}.
\end{align}
The symbol $B_\mu$ denotes the spin 3/2 Rarita-Schwinger field, which
is subject to the constraints 
\begin{align}
  p^\mu B_\mu = 0 \hspace{0.5 cm}{\rm and}\hspace{0.5 cm}
  \gamma^\mu B_\mu = 0 .
\end{align}
Within this effective Lagrangian, the coupling constants are defined
as follows~\cite{Liu:2011xc,Chen:2019asm}: $\beta_{\bar{3}} = 6/g_V$,
$\beta_6=-2\beta_{\bar{3}}$, $\lambda_6=-3.31\,\mathrm{GeV}^{-1}$,
$\lambda_{6\bar{3}}=-\lambda_6/\sqrt{8}$, $g_1=0.942$ and
$g_4=0.999$, $l_{\bar{3}}= -3.1$ and $l_6=-2l_{\bar{3}}$. 
The sign conventions we implement are in line with those 
established in Refs.~\cite{Chen:2019asm, Dong:2021juy}.  

For hidden-charm channels, we concentrate solely on vector charmonia 
due to their immediate experimental significance. Nevertheless,
including pseudoscalar charmonium states is straightforward,
as we extend heavy quark spin symmetry principles to the charmonium
domain as well~\cite{Casalbuoni:1992fd}. The interaction between heavy
mesons and charmonium is governed by the following effective
Lagrangian~\cite{Colangelo:2003sa} 
\begin{align}
    \mathcal{L}_{PPJ/\psi} &= -ig_\psi M\sqrt{m_{J}}\left(J/\psi^\mu P^\dagger\overleftrightarrow{\partial_\mu}P{'}^{\dagger}\right) + \mathrm{h.c.,}\\
    \mathcal{L}_{P^*PJ/\psi} &= ig_\psi\sqrt{\frac{MM^*}{m_{J}}}\epsilon^{\mu\nu\alpha\beta} \partial_\mu J/\psi_\nu\left(P^\dagger\overleftrightarrow{\partial_\alpha}P^*{'}^\dagger_\beta+P_{\beta}^{*\dagger}\overleftrightarrow{\partial_\alpha}P{'}^{\dagger}\right)+\mathrm{h.c.,}\\
    \mathcal{L}_{P^*P^*J/\psi} &= ig_\psi M^*\sqrt{m_J}(g^{\mu\nu}g^{\alpha\beta}-g^{\mu\alpha}g^{\nu\beta}+g^{\mu\beta}g^{\nu\alpha}) \left(J/\psi_\mu P_{\nu}^{*\dagger}\overleftrightarrow{\partial_\alpha}P^*{'}^\dagger_\beta\right)+\mathrm{h.c.}.
\end{align}
Given the lack of experimental measurements for the $J/\psi \to
D\bar{D}$ decay, Shimizu et al.~\cite{Shimizu:2017xrg} developed an
estimation method for the coupling constant $g_\psi$: they initially
calculated the coupling constant $g_{\phi K\bar{K}}$ from the
experimental decay width of $\phi \to K\bar{K}$. Assuming that the
$J/\psi$ decay mechanisms parallel to those of the $\phi$, with
differences attributed mainly to mass variations, they derived an
estimated coupling constant value of $g_\psi =
0.679\,\mathrm{GeV}^{-3/2}$. 

Following Ref.~\cite{Shimizu:2017xrg}, we obtain the coupling
constants between heavy baryons and heavy mesons as: 
\begin{align}
    \mathcal{L}_{B_8B_3P} &=
   g_{I\bar{3}}\sqrt{M}\bar{B}_{8}\gamma_5
   \left(P' B_{\bar{3}}\right)_8+\mathrm{h.c.},\\ 
    \mathcal{L}_{B_8B_3P^*} &=
    g_{I\bar{3}}\sqrt{M^*}\bar{B}_{8}\gamma^\mu
    \left(P'^*_\mu B_{\bar{3}}\right)_8 +\mathrm{h.c.},\\ 
    \mathcal{L}_{B_8B_6P}&= -g_{I6} \sqrt{3M} \bar{B}_{8}\gamma_5 \left(P' B_{6}\right)_8
        + \mathrm{h.c.},\\ 
    \mathcal{L}_{B_8B_6P^*}&= g_{I6}\sqrt{\frac{M^*}{3}}
   \bar{B}_{8}\gamma^\mu \left(P'^*_\mu B_{6}\right)_8 +
   \mathrm{h.c.},\\ 
    \mathcal{L}_{B_8B_6^*P^*}&= -2 g_{I6}\sqrt{M^*}
    \bar{B}_{8}^\mu\gamma_5 \left(P'^*_\mu B_{6}\right)_8 +
    \mathrm{h.c.}
\end{align}
The octet baryon matrix is expressed as
\begin{align}
    B_8 = 
    \begin{pmatrix}
        \frac{1}{\sqrt{2}} \Sigma^0+\frac{1}{\sqrt{6}}\Lambda & \Sigma^+ & p\\
        \Sigma^- & -\frac{1}{\sqrt{2}} \Sigma^0+\frac{1}{\sqrt{6}}\Lambda & n\\
        \Xi^- & \Xi^0 & -\frac{2}{\sqrt{6}}\Lambda
    \end{pmatrix},
\end{align}
while the octet components of the tensor product between triplet charmed meson and singly-heavy baryon are expressed as
\begin{align}
   \left(P' B_{\bar{3}}\right)_8 
   &= 
   \begin{pmatrix}
      \frac{2}{3}\bar{D}^0\Xi_c^0 + \frac{1}{3}D^-\Xi_c^+ - \frac{1}{3}D_s^-\Lambda_c^+ & -\bar{D}^0\Xi_c^{+} & \bar{D}^0\Lambda_c^{+} \\[0.5em]
      D^-\Xi_c^{0} & -\frac{2}{3}D^-\Xi_c^+ -\frac{1}{3}\bar{D}^0\Xi_c^0 - \frac{1}{3}D_s^-\Lambda_c^+ & D^-\Lambda_c^{+} \\[0.5em]
      D_s^-\Xi_c^{0} & -D_s^-\Xi_c^{+} & \frac{2}{3}D_s^-\Lambda_c^+ -\frac{1}{3}\bar{D}^0\Xi_c^0 +\frac{1}{3}D^-\Xi_c^+
   \end{pmatrix},\\
   \left(P' B_{6}\right)_8 &= 
   \begin{pmatrix}
      \frac{1}{\sqrt{2}}(D_s^-\Sigma_c^+-D^-\Xi'^+_c) & \frac{1}{\sqrt{2}}\bar{D}^0\Xi'^+_c -D_s^-\Sigma_c^{++} & D^-\Sigma_c^{++}-\frac{1}{\sqrt{2}}\bar{D}^0\Sigma^+_c \\[0.5em]
      D_s^-\Sigma_c^{0}-\frac{1}{\sqrt{2}}D^-\Xi'^0_c & \frac{1}{\sqrt{2}}(\bar{D}^0\Xi'^0_c-D_s^-\Sigma_c^+) & \frac{1}{\sqrt{2}}D^-\Sigma_c^{+}-\bar{D}^0\Sigma_c^0 \\[0.5em]
      \frac{1}{\sqrt{2}}D_s^-\Xi'^0_c-D^-\Omega_c^0 & \bar{D}^0\Omega_c^0-\frac{1}{\sqrt{2}}D_s^-\Xi'^+_c & \frac{1}{\sqrt{2}}(D^-\Xi'^+_c-\bar{D}^0\Xi'^0_c)
   \end{pmatrix}.
\end{align}
The structure is similar when we deal with charmed vector mesons or
$J=3/2$ sextet baryons. In this way, we have an SU(3) symmetric
effective Lagrangian for the transition of singly heavy baryon to
light baryon through charmed meson emission. From
Ref.~\cite{Shimizu:2017xrg}, we implement the coupling constants
$g_{I\bar{3}} = -9.88\,\mathrm{GeV}^{-1/2}$ and $g_{I6} =
1.14\,\mathrm{GeV}^{-1/2}$. It is worth noting that the coupling to
hidden-charm channels plays only a minor role in the resonance
production mechanism. The predicted masses of hidden-charm pentaquarks
show little variation with changes in these coupling constants. 

\begin{table}[htbp]
  \caption{\label{tab:1}Values of the IS factors 
          for the corresponding $t$-channel diagrams for
          the given reactions.  
        } 
   \renewcommand{\arraystretch}{1.2}
  \begin{ruledtabular}
  \centering\begin{tabular}{lcr}
   \multirow{2}{*}{Reactions} & \multirow{2}{*}{Exchange particles} & 
   \multirow{2}{*}{IS} \\
   & &
   \\
   \hline\\[-2.5ex]
     $J/\psi\Xi\to\bar{D}_s\Xi_c$ 
     & $\bar{D}_s$, $\bar{D}_s^*$  & $-1$ \\
     $J/\psi\Xi\to\bar{D}_s\Xi_c'$ 
     & $\bar{D}_s$, $\bar{D}_s^*$  & $\frac{1}{2}\sqrt{2}$ \\
     $J/\psi\Xi\to\bar{D}\Omega_c$ 
     & $\bar{D}$, $\bar{D}^*$  & $-1$ \\
     $J/\psi\Xi\to\bar{D}_s^*\Xi_c$ 
     & $\bar{D}_s$, $\bar{D}_s^*$  & $-1$ \\
     $J/\psi\Xi\to\bar{D}_s\Xi_c^*$ 
     & $\bar{D}_s$, $\bar{D}_s^*$  & $\frac{1}{2}\sqrt{2}$ \\
     $J/\psi\Xi\to\bar{D}\Omega_c^*$ 
     & $\bar{D}$, $\bar{D}^*$ & $-1$ \\
     $J/\psi\Xi\to\bar{D}_s^*\Xi_c'$ 
     & $\bar{D}_s$, $\bar{D}_s^*$  & $\frac{1}{2}\sqrt{2}$ \\
     $J/\psi\Xi\to\bar{D}^*\Omega_c$ 
     & $\bar{D}$, $\bar{D}^*$ & $-1$ \\
     $J/\psi\Xi\to\bar{D}_s^*\Xi_c^*$ 
     & $\bar{D}_s$, $\bar{D}_s^*$  & $\frac{1}{2}\sqrt{2}$ \\
     $J/\psi\Xi\to\bar{D}^*\Omega_c^*$ 
     & $\bar{D}$, $\bar{D}^*$ & $-1$ \\
     $\bar{D}_s\Xi_c\to\bar{D}_s\Xi_c$ 
     & $\phi$  & $1$ \\
     & $\sigma$  & $2$ \\
     $\bar{D}_s\Xi_c\to\bar{D}_s\Xi_c'$ 
     & $\phi$  & $-\frac{1}{2}\sqrt{2}$ \\
     $\bar{D}_s\Xi_c\to\bar{D}\Omega_c$ 
     & $K^*$  & $1$ \\
     $\bar{D}_s\Xi_c\to\bar{D}_s^*\Xi_c$ 
     & $\phi$  & $1$ \\
     $\bar{D}_s\Xi_c\to\bar{D}_s\Xi_c^*$ 
     & $\phi$  & $-\frac{1}{2}\sqrt{2}$ \\
     $\bar{D}_s\Xi_c\to\bar{D}\Omega_c^*$ 
     & $K^*$ & $1$ \\
     $\bar{D}_s\Xi_c\to\bar{D}_s^*\Xi_c'$ 
     & $\eta$  & $-\frac{1}{2}\sqrt{2}$ \\
     & $\phi$  & $-\frac{1}{2}\sqrt{2}$ \\
     $\bar{D}_s\Xi_c\to\bar{D}^*\Omega_c$ 
     & $K$, $K^*$ & $1$ \\
     $\bar{D}_s\Xi_c\to\bar{D}_s^*\Xi_c^*$ 
     & $\eta$  & $-\frac{1}{2}\sqrt{2}$ \\
     & $\phi$  & $-\frac{1}{2}\sqrt{2}$ \\
     $\bar{D}_s\Xi_c\to\bar{D}^*\Omega_c^*$ 
     & $K$, $K^*$ & $1$ \\
     $\bar{D}_s\Xi_c'\to\bar{D}_s\Xi_c'$ 
     & $\phi$  & $\frac{1}{2}$ \\
     & $\sigma$  & $1$ \\
     $\bar{D}_s\Xi_c'\to\bar{D}\Omega_c$ 
     & $K^*$ & $\frac{1}{2}\sqrt{2}$ \\
     $\bar{D}_s\Xi_c'\to\bar{D}_s^*\Xi_c$ 
     & $\eta$  & $-\frac{1}{2}\sqrt{2}$ \\
     & $\phi$  & $-\frac{1}{2}\sqrt{2}$ \\
     $\bar{D}_s\Xi_c'\to\bar{D}_s\Xi_c^*$ 
     & $\phi$  & $\frac{1}{2}$ \\
     $\bar{D}_s\Xi_c'\to\bar{D}\Omega_c^*$ 
     & $K^*$ & $\frac{1}{2}\sqrt{2}$ \\
     $\bar{D}_s\Xi_c'\to\bar{D}_s^*\Xi_c'$ 
     & $\eta$  & $\frac{1}{6}$ \\
     & $\phi$  & $\frac{1}{2}$ \\
     $\bar{D}_s\Xi_c'\to\bar{D}^*\Omega_c$ 
     & $K$, $K^*$ & $\frac{1}{2}\sqrt{2}$ \\
     $\bar{D}_s\Xi_c'\to\bar{D}_s^*\Xi_c^*$ 
     & $\eta$  & $\frac{1}{6}$ \\
     & $\phi$  & $\frac{1}{2}$ \\
     $\bar{D}_s\Xi_c'\to\bar{D}^*\Omega_c^*$ 
     & $K$, $K^*$ & $\frac{1}{2}\sqrt{2}$ \\
     $\bar{D}\Omega_c\to\bar{D}\Omega_c$ 
     & $\sigma$  & $1$ \\
     $\bar{D}\Omega_c\to\bar{D}_s^*\Xi_c$ 
     & $\bar{K}$, $\bar{K}^*$ & $1$ \\
     $\bar{D}\Omega_c\to\bar{D}_s\Xi_c^*$ 
     & $\bar{K}^*$ & $\frac{1}{2}\sqrt{2}$ \\
%
%
     $\bar{D}\Omega_c\to\bar{D}_s^*\Xi_c'$ 
     & $\bar{K}$, $\bar{K}^*$ & $\frac{1}{2}\sqrt{2}$ \\
     $\bar{D}\Omega_c\to\bar{D}^*\Omega_c$ 
     & $\eta$ & $-\frac{1}{6}\sqrt{2}$ \\
     $\bar{D}\Omega_c\to\bar{D}_s^*\Xi_c^*$ 
     & $\bar{K}$, $\bar{K}^*$ & $\frac{1}{2}\sqrt{2}$ \\
     $\bar{D}\Omega_c\to\bar{D}^*\Omega_c^*$ 
     & $\eta$ & $-\frac{1}{6}\sqrt{2}$ \\
  \end{tabular}
    \end{ruledtabular}
\end{table}
\begin{table}[htbp]
  \addtocounter{table}{-1}
  \caption{Values of the IS factors
          for the corresponding $t$-channel diagrams for
          the given reactions (continued).  
        } 
   \renewcommand{\arraystretch}{1.2}
  \begin{ruledtabular}
  \centering\begin{tabular}{lcr}
   \multirow{2}{*}{Reactions} & \multirow{2}{*}{Exchange particles} & 
   \multirow{2}{*}{IS} \\
   & &
   \\
   \hline\\[-2.5ex]
     $\bar{D}_s^*\Xi_c\to\bar{D}_s^*\Xi_c$ 
     & $\phi$  & $1$ \\
     & $\sigma$  & $2$ \\
     $\bar{D}_s^*\Xi_c\to\bar{D}_s\Xi_c^*$ 
     & $\eta$ & $-\frac{1}{2}\sqrt{2}$ \\
     & $\phi$  & $-\frac{1}{2}\sqrt{2}$ \\
     $\bar{D}_s^*\Xi_c\to\bar{D}\Omega_c^*$ 
     & $K$, $K^*$ & $1$ \\
     $\bar{D}_s^*\Xi_c\to\bar{D}_s^*\Xi_c'$ 
     & $\eta$  & $-\frac{1}{2}\sqrt{2}$ \\
     & $\phi$  & $-\frac{1}{2}\sqrt{2}$ \\
     $\bar{D}_s^*\Xi_c\to\bar{D}^*\Omega_c$ 
     & $K$, $K^*$ & $1$ \\
     $\bar{D}_s^*\Xi_c\to\bar{D}_s^*\Xi_c^*$ 
     & $\eta$  & $-\frac{1}{2}\sqrt{2}$ \\
     & $\phi$  & $-\frac{1}{2}\sqrt{2}$ \\
     $\bar{D}_s^*\Xi_c\to\bar{D}^*\Omega_c^*$ 
     & $K$, $K^*$ & $1$ \\
     $\bar{D}_s\Xi_c^*\to\bar{D}_s\Xi_c^*$ 
     & $\phi$  & $\frac{1}{2}$ \\
     & $\sigma$  & $1$ \\
     $\bar{D}_s\Xi_c^*\to\bar{D}\Omega_c^*$ 
     & $K^*$ & $\frac{1}{2}\sqrt{2}$ \\
     $\bar{D}_s\Xi_c^*\to\bar{D}_s^*\Xi_c'$ 
     & $\eta$  & $\frac{1}{6}$ \\
     & $\phi$  & $\frac{1}{2}$ \\
     $\bar{D}_s\Xi_c^*\to\bar{D}^*\Omega_c$ 
     & $K$, $K^*$ & $\frac{1}{2}\sqrt{2}$ \\
     $\bar{D}_s\Xi_c^*\to\bar{D}_s^*\Xi_c^*$ 
     & $\eta$  & $\frac{1}{6}$ \\
     & $\phi$  & $\frac{1}{2}$ \\
     $\bar{D}_s\Xi_c^*\to\bar{D}^*\Omega_c^*$ 
     & $K$, $K^*$ & $\frac{1}{2}\sqrt{2}$ \\
     $\bar{D}\Omega_c^*\to\bar{D}\Omega_c^*$ 
     & $\sigma$ & $1$ \\
     $\bar{D}\Omega_c^*\to\bar{D}_s^*\Xi_c'$ 
     & $\bar{K}$, $\bar{K}^*$ & $\frac{1}{2}\sqrt{2}$ \\
     $\bar{D}\Omega_c^*\to\bar{D}^*\Omega_c$ 
     & $\eta$ & $-\frac{1}{6}\sqrt{2}$ \\
     $\bar{D}\Omega_c^*\to\bar{D}_s^*\Xi_c^*$ 
     & $\bar{K}$, $\bar{K}^*$ & $\frac{1}{2}\sqrt{2}$ \\
     $\bar{D}\Omega_c^*\to\bar{D}^*\Omega_c^*$ 
     & $\eta$ & $-\frac{1}{6}\sqrt{2}$ \\
     $\bar{D}_s^*\Xi_c'\to\bar{D}_s^*\Xi_c'$ 
     & $\eta$  & $\frac{1}{6}$ \\
     & $\phi$  & $\frac{1}{2}$ \\
     & $\sigma$  & $1$ \\
     $\bar{D}_s^*\Xi_c'\to\bar{D}^*\Omega_c$ 
     & $K$, $K^*$ & $\frac{1}{2}\sqrt{2}$ \\
     $\bar{D}_s^*\Xi_c'\to\bar{D}_s^*\Xi_c^*$ 
     & $\eta$  & $\frac{1}{6}$ \\
     & $\phi$  & $\frac{1}{2}$ \\
     $\bar{D}_s^*\Xi_c'\to\bar{D}^*\Omega_c^*$ 
     & $K$, $K^*$ & $\frac{1}{2}\sqrt{2}$ \\
     $\bar{D}^*\Omega_c\to\bar{D}^*\Omega_c$ 
     & $\eta$ & $-\frac{1}{6}\sqrt{2}$ \\
     & $\sigma$ & $1$ \\
     $\bar{D}^*\Omega_c\to\bar{D}_s^*\Xi_c^*$ 
     & $\bar{K}$, $\bar{K}^*$ & $\frac{1}{2}\sqrt{2}$ \\
     $\bar{D}^*\Omega_c\to\bar{D}^*\Omega_c^*$ 
     & $\eta$ & $-\frac{1}{6}\sqrt{2}$ \\
     $\bar{D}_s^*\Xi_c^*\to\bar{D}_s^*\Xi_c^*$ 
     & $\eta$  & $\frac{1}{6}$ \\
     & $\phi$  & $\frac{1}{2}$ \\
     & $\sigma$ & $1$ \\
     $\bar{D}_s^*\Xi_c^*\to\bar{D}^*\Omega_c^*$ 
     & $K$, $K^*$ & $\frac{1}{2}\sqrt{2}$ \\
     $\bar{D}^*\Omega_c^*\to\bar{D}^*\Omega_c^*$ 
     & $\eta$ & $-\frac{1}{6}\sqrt{2}$ \\
     & $\sigma$ & $1$ \\
  \end{tabular}
    \end{ruledtabular}
\end{table}

The Feynman amplitude for a one-meson exchange diagram can be
expressed as   
\begin{align}
  \mathcal{A}_{\lambda'_1\lambda'_2,\lambda_1\lambda_2}
  = \mathrm{IS} \,F^2(q^2)\,\Gamma_{\lambda'_1\lambda'_2}(p'_1,p'_2)
  \mathcal P(q)\Gamma_{\lambda_1\lambda_2}(p_1,p_2) ,
\end{align}
where $\lambda_i$ and $p_i$ denote respectively the helicity and momentum of
each particle participating in the process, while $q$ designates the
momentum transfer. The IS factor contains the SU(3)
Clebsch-Gordan coefficient and isospin factor, with values for each
exchange diagram documented in Table~\ref{tab:1}. We derive the vertex
$\Gamma$ from the effective Lagrangian, and express the propagators
for spin-0 and spin-1 mesons as   
\begin{align}
  \mathcal{P}(q) &= \frac{1}{q^2-m^2},\;\;\;
  \mathcal{P}_{\mu\nu}(q) = \frac{1}{q^2-m^2}
  \left(-g_{\mu\nu}+\frac{q_\mu q_\nu}{m^2}\right).
\end{align}
For the heavy-meson propagators, we adopt the same form as for the
light mesons, since the heavy-quark mass is finite. 

Since hadrons have finite hadron sizes, we need to introduce a form
factor at each vertex. We employ the following form of the form
factor~\cite{Kim:1994ce}    
\begin{align}
  F(q^2) = \left(\frac{n\Lambda^2-m^2}
  {n\Lambda^2-q^2}
  \right)^n,
\label{eq:13}
\end{align}
where the value of $n$ depends on the momentum power present
in the vertex. A key benefit of this parametrization is that modifying
$\Lambda$ becomes unnecessary when $n$ changes. It should be observed
that as $n$ approaches infinity, Eq.~\eqref{eq:13} transforms into a
Gaussian form. While it is difficult to determine the cutoff masses
$\Lambda$ in Eq.~\eqref{eq:13} on account of a lack of experimental
information, we utilize a physical feature of hadron sizes
to reduce the uncertainties due to the cutoff masses. Previous
studies indicate that heavy hadrons exhibit more compact sizes than
their lighter counterparts~\cite{Kim:2018nqf, Kim:2021xpp}, suggesting
that higher cutoff masses are appropriate for heavy hadrons compared
to light hadrons. Based on this reasoning, we define a 
value of the reduced cutoff mass as $\Lambda_0 := \Lambda -
m$, where $m$ corresponds to the mass of the exchange particle. In the
absence of experimental data on the $P_{c\bar{c}ss}$, we apply a
consistent value of 700 MeV across all reduced cutoff parameters
$\Lambda_0$.    

In the context of the strong interaction, we can use parity
invariance to reduce the number of independent amplitudes
required. The parity relation is mathematically formulated as: 
\begin{align}
  \mathcal{A}_{-\lambda'_1-\lambda'_2,-\lambda_1-\lambda_2} =
  \eta(\eta')^{-1}\,
  \mathcal{A}_{\lambda'_1\lambda'_2,\lambda_1\lambda_2},  
  \label{eq:ampi}
\end{align}
with $\eta$ and $\eta'$ defined by:
\begin{align}
    \eta = \eta_1\eta_2(-1)^{J-s_1-s_2}, \hspace{0.5 cm} \eta' =
  \eta_1'\eta_2'(-1)^{J-s_1'-s_2'}. 
\end{align}
In these expressions, $\eta_i$ and $s_i$ represent the intrinsic
parity and spin of the particle respectively, while $J$ corresponds to
the total angular momentum. This relationship significantly decreases
the computational time required for numerical calculations. 

To enhance computational efficiency and elucidate the spin-parity
classifications of $P_{c\bar{c}ss}$ states, we implement a
partial-wave decomposition of the $\mathcal{V}$ and $\mathcal{T}$
matrices. This approach yields a simplified one-dimensional integral
equation:  
\begin{align}
  \mathcal{T}^{J(fi)}_{\lambda'\lambda} (\mathrm{p}',\mathrm{p}) = 
  \mathcal{V}^{J(fi)}_{
  \lambda'\lambda} (\mathrm{p}',\mathrm{p}) + \frac{1}{(2\pi)^3}
  \sum_{k,\lambda_k}\int
  \frac{\mathrm{q}^2d\mathrm{q}}{2E_{k1}E_{k2}}
  \mathcal{V}^{J(fk)}_{\lambda'\lambda_k}(\mathrm{p}',
  \mathrm{q})\frac{E_k}{
  s-E_k^2+i\varepsilon} \mathcal{T}^{J(ki)}_{\lambda_k\lambda}
  (\mathrm{q},\mathrm{p}),
  \label{eq:BS-1d}
\end{align}
where the helicities of the final, initial, and intermediate states
are denoted by $\lambda'=\{\lambda'_1,\lambda'_2\}$,
$\lambda=\{\lambda_1,\lambda_2\}$, and
$\lambda_k=\{\lambda_{k1},\lambda_{k2}\}$, respectively. The variables
$\mathrm{p}'$, $\mathrm{p}$, and $\mathrm{q}$ correspond to the
magnitudes of the momentum vectors $\bm{p}'$, $\bm{p}$, and
$\bm{q}$. We express the partial-wave kernel amplitudes
$\mathcal{V}_{\lambda'\lambda}^{J(fi)}$ as  
\begin{equation}
  \mathcal{V}^{J(fi)}_{\lambda'\lambda}(\mathrm{p}',\mathrm{p}) = 
  2\pi \int d( \cos\theta) \,
        d^{J}_{\lambda_1-\lambda_2,\lambda'_1-\lambda'_2}(\theta)\,
        \mathcal{V}^{fi}_{\lambda'\lambda}(\mathrm{p}',\mathrm{p},\theta),
  \label{eq:pwd}
\end{equation} 
where $\theta$ represents the scattering angle and
$d^{J}_{\lambda\lambda'}(\theta)$ refers to the reduced Wigner $D$
functions. 

The integral equation presented in Eq.~\eqref{eq:BS-1d} contains a
singularity coming from the two-body propagator $\mathcal{G}$. To
effectively manage this singularity, we treat its
singular component independently. The resulting regularized integral
equation takes the form: 
\begin{align}
  \mathcal{T}^{fi}_{\lambda'\lambda} (\mathrm{p}',\mathrm{p}) = 
  \mathcal{V}^{fi}_{
  \lambda'\lambda} (\mathrm{p}',\mathrm{p}) + \frac{1}{(2\pi)^3}
  \sum_{k,\lambda_k}\left[\int_0^{\infty}d\mathrm{q}
  \frac{\mathrm{q}E_k}{E_{k1}E_{k2}}\frac{\mathcal{F}(\mathrm{q})
  -\mathcal{F}(\tilde{\mathrm{q}}_k)}{s-E_k^2}+ \frac{1}{2\sqrt{s}}
  \left(\ln\left|\frac{\sqrt{s}-E_k^{\mathrm{thr}}}{\sqrt{s}
  +E_k^{\mathrm{thr}}}\right|-i\pi\right)\mathcal{F}
  (\tilde{\mathrm{q}}_k)\right],
  \label{eq:BS-1d-reg}
\end{align}
where
\begin{align}
  \mathcal{F}(\mathrm{q})=\frac{1}{2}\mathrm{q}\,
  \mathcal{V}^{fk}_{\lambda'\lambda_k}(\mathrm{p}',
  \mathrm{q})\mathcal{T}^{ki}_{\lambda_k\lambda}(\mathrm{q},\mathrm{p}) ,
\end{align}
and $\tilde{\mathrm{q}}_k$ represents the momentum $\mathrm{q}$ when
$E_{k1}+E_{k2}=\sqrt{s}$. This regularization procedure is implemented
exclusively when the total energy $\sqrt{s}$ is greater than the
threshold energy of the $k$-th channel $E_k^{\mathrm{thr}}$. Notably,
the form factors incorporated in the kernel amplitudes $\mathcal{V}$
guarantee the unitarity of the transition amplitudes in the
high-momentum domain. 

For the numerical evaluation of the $\mathcal{T}$ matrix in
Eq.~\eqref{eq:BS-1d-reg}, we decompose the $\mathcal{V}$ matrix in the
helicity basis and represent it in momentum space, with momenta
determined via the Gaussian quadrature method. The $\mathcal{T}$
matrix is subsequently derived through the Haftel–Tabakin matrix
inversion technique~\cite{Haftel:1970zz}: 
\begin{align}
  \mathcal{T} = \left(1-\mathcal{V}\tilde{\mathcal{G}}\right)^{-1} 
  \mathcal{V}.
\end{align}
The derived $\mathcal{T}$ matrix, expressed in the helicity basis,
lacks definite parity. To investigate the parity classifications of
the $P_{c\bar{c}ss}$ states, we transform the transition amplitudes
into partial-wave amplitudes with well-defined parity: 
\begin{align}
  \mathcal{T}^{J\pm}_{\lambda'\lambda} =
  \frac{1}{2}\left[\mathcal{T}^{J}_{\lambda'\lambda} \pm
  \eta_1\eta_2(-1)^{s_1+s_2+\frac{1}{2}}
  \mathcal{T}^{J}_{\lambda'-\lambda}\right], 
\end{align}
where $\mathcal{T}^{J\pm}$ represents the partial-wave transition
amplitude with total angular momentum $J$ and parity $(-1)^{J\pm
  1/2}$. The coefficient $1/2$ ensures that no additional scaling
factor is required when reverting to the partial-wave component: 
\begin{align}
    \mathcal{T}^{J}_{\lambda'\lambda} =
  \mathcal{T}^{J+}_{\lambda'\lambda} +\mathcal{T}^{J-}_{\lambda'\lambda}.
\end{align}
It is important to note that decomposing the partial-wave component
with definite parity in Eq.~\eqref{eq:BS-1d} is superfluous, as parity
invariance is already integrated into both the effective Lagrangian
and the amplitude calculations, as demonstrated in
Eq.~\eqref{eq:ampi}.   

For our analysis of dynamically generated resonances, we reformulate
the $\mathcal{T}$ matrix in the $IJL$ particle
basis~\cite{Machleidt:1987hj}. The transformational relationships
between the $\mathcal{T}$ matrix elements across these two bases are
expressed as: 
\begin{align}
  \mathcal{T}^{JS'S}_{L'L} = \frac{\sqrt{(2L+1)(2L'+1)}}{2J+1}
  \sum_{\lambda'_1\lambda'_2\lambda_1\lambda_2}
  \left(L'0S'\lambda'|J\lambda'\right)
  \left(s'_1\lambda'_1s'_2-\lambda'_2|S'\lambda'\right)
  \left(L0S\lambda|J\lambda\right)
  \left(s_1\lambda_1s_2-\lambda_2|S\lambda\right)
  \mathcal{T}^{J}_{\lambda'_1\lambda'_2,\lambda_1\lambda_2} .
\end{align}
In the present study, we focus exclusively on the diagonal component
$\mathcal{T}^{JS}_{L}$, as it has the most significant implications
for the production of the double-strangeness hidden-charm pentaquarks. 

\section{Results and discussions \label{sec:3}}
As mentioned previously, we have introduced the eleven different
channels to investigate the dynamical generation of the
double-strangeness hidden-charm pentaquark states. It implies that we
also have eleven different two-particle thresholds: $J/\psi \Xi$,
$\bar{D}_s\Xi_c$, $\bar{D}_s \Xi_c'$, $\bar{D}\Omega_c$,
$\bar{D}_s^*\Xi_c$, $\bar{D}_s \Xi_c^*$, $\bar{D}\Omega_c^*$,
$\bar{D}_s^* \Xi_c'$, $\bar{D}^*\Omega_c$, $\bar{D}_s^* \Xi_c^*$, and
$\bar{D}^*\Omega_c^*$ in the order of increasing the threshold
energies.  In the previous works on the hidden-charm pentaquark states
with $S=0$~\cite{Clymton:2024fbf} and $S=-1$~\cite{Clymton:2025hez},
those with negative parity were always located below the thresholds,
whereas those with positive parity were found above the thresholds. As
will be shown in this work, we will see the same tendency in the case
of the double-strangeness hidden-charm pentaquarks. 
We limit our discussion to the cases where total isospin equals
$1/2$, anticipating that the $P_{c\bar{c}ss}$ may be observed in the
$J/\psi \Xi$ invariant mass distributions. Since experimental data are 
not yet available, we will focus on finding most probable pole
position of the $P_{c\bar{c}ss}$ resonances. Table~\ref{tab:1} lists
all the parameters used in the exchange diagram to generate the kernel
Feynman amplitudes. Regarding theoretical uncertainties arising from
the cutoff masses, we will discuss it at the end of this section.

\subsection{Negative parity}

\begin{figure}[htbp]
  \centering
  \includegraphics[scale=0.51]{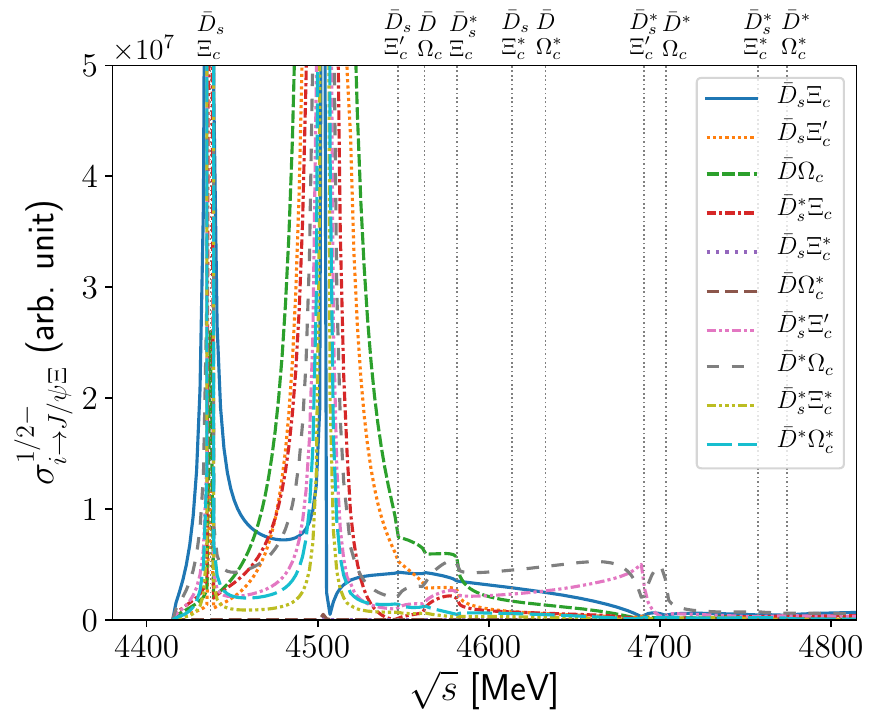}
  \includegraphics[scale=0.51]{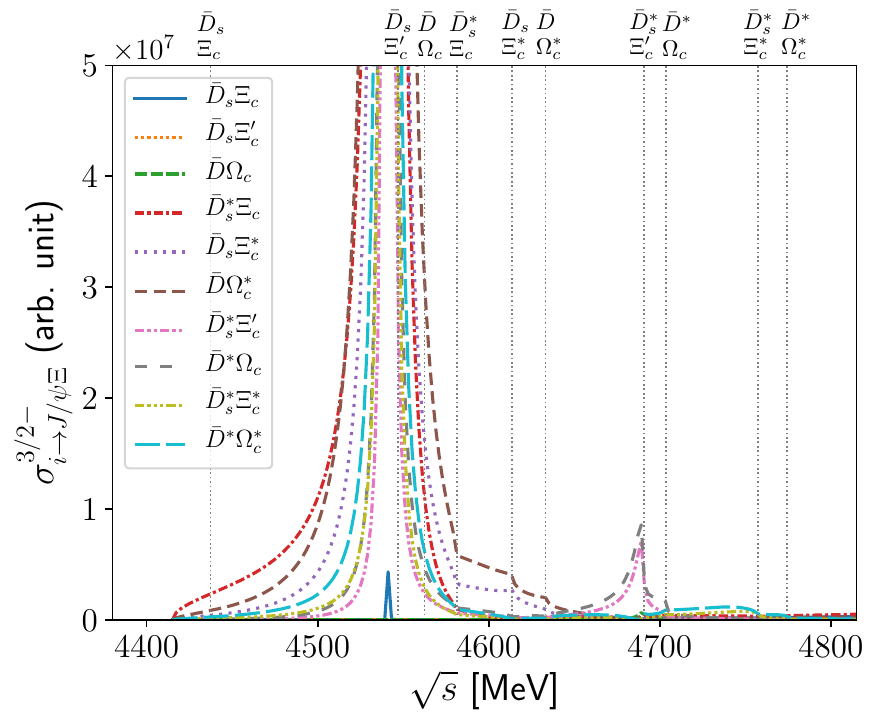}
  \includegraphics[scale=0.51]{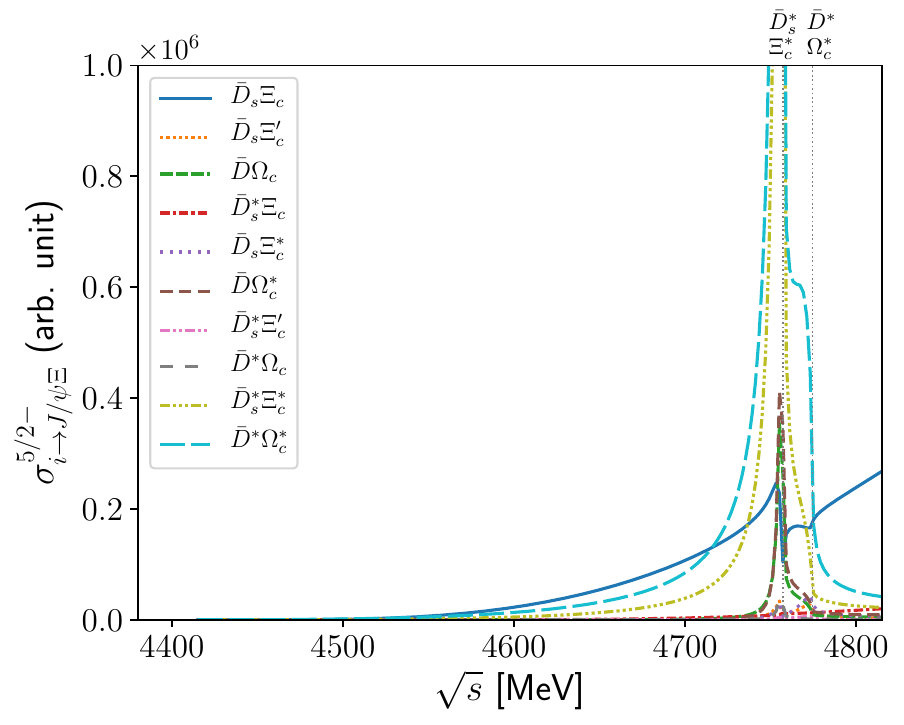}
  \caption{$i\to J/\psi \Xi$ transition partial-wave cross sections
    for the given total angular momenta $J=1/2,\,3/2,\, 5/2$ with
    negative parity, which correspond to the spins and parity of
    $P_{c\bar{c}ss}$, as functions of the total energy. $i$ denotes
    the initial two-particle states, as expressed in the legend.}   
  \label{fig:4} 
\end{figure}
Figure~\ref{fig:4} draws the results for the $i\to J/\psi \Xi$
transition partial-wave cross sections for the total angular momentum
$J=1/2,\,3/2,\, 5/2$ with negative parity, which correspond to the
spins and parity of $P_{c\bar{c}ss}$, as functions of the total
energy. $i$ represents the initial two-particle states, as expressed
in the legend of Fig.~\ref{fig:4}. As shown in the upper-left panel of 
Fig.~\ref{fig:4}, we identify three resonances with spin-parity
$J^P=1/2^-$. On the other hand, the upper-right panel and lower panel
of Fig.~\ref{fig:4} demonstrate a resonance with $J^P=3/2^-$ and that
with $J^P=5/2^-$, respectively. Table~\ref{tab:2} lists the pole
positions of these five negative-parity pentaquark resonances in the
complex plane, and numerical results for the coupling strengths to the
coupled channels involved.  

  \begin{table}
  \caption{\label{tab:2}Coupling strengths of the five $P_{c\bar{c}ss}$'s
      with $J^P=1/2^-$, $3/2^-$, and $5/2^-$.}  
   \centering
   \begin{tabular*}{\linewidth}{@{\extracolsep{\fill}} lccccc}
      \toprule
    $J^P$ & \multicolumn{3}{c}{$1/2^-$} & \multicolumn{1}{c}{$3/2^-$} & \multicolumn{1}{c}{$5/2^-$} \\
    & $P_{c\bar{c}ss}(4437)$ & $P_{c\bar{c}ss}(4504)$ & $P_{c\bar{c}ss}(4704)$ & $P_{c\bar{c}ss}(4541)$ & $P_{c\bar{c}ss}(4757)$ \\
    $\sqrt{s_R}$ [MeV] & $4437.2-i0.002$ & $4504.1-i0.2$ & $4703.7-i10.6$ & $4541.3-i0.04$ & $4756.5-i1.7$ \\
   \midrule
   $g_{J/\psi\Xi({}^2S_J)}$          & $0.10+i0.00$ & $0.29+i0.02$ & $0.03+i0.11$     & $-$ & $-$ \\
   $g_{J/\psi\Xi({}^2D_J)}$          & $-$ & $-$ & $-$                                & $0.01-i0.00$ & $0.01+i0.00$ \\
   $g_{J/\psi\Xi({}^4S_J)}$          & $-$ & $-$ & $-$                                & $0.27+i0.00$ & $-$ \\
   $g_{J/\psi\Xi({}^4D_J)}$          & $0.00+i0.00$ & $0.01-i0.00$ & $0.01+i0.02$     & $0.04+i0.00$ & $0.02+i0.00$ \\
   $g_{\bar{D}_s\Xi_c({}^2S_J)}$     & $4.60+i0.01$ & $-0.03+i0.67$ & $0.85+i0.44$    & $-$ & $-$ \\
   $g_{\bar{D}_s\Xi_c({}^2D_J)}$     & $-$ & $-$ & $-$                                & $0.03+i0.00$ & $-0.24-i0.17$ \\
   $g_{\bar{D}_s\Xi_c'({}^2S_J)}$    & $0.17+i0.00$ & $11.72-i0.01$ & $-2.70-i0.58$   & $-$ & $-$ \\
   $g_{\bar{D}_s\Xi_c'({}^2D_J)}$    & $-$ & $-$ & $-$                                & $0.00+i0.00$ & $-0.21-i0.15$ \\
   $g_{\bar{D}\Omega_c({}^2S_J)}$    & $-0.23-i0.00$ & $-15.86+i0.02$ & $0.08-i2.24$  & $-$ & $-$ \\
   $g_{\bar{D}\Omega_c({}^2D_J)}$    & $-$ & $-$ & $-$                                & $-0.00-i0.00$ & $0.95+i0.25$ \\
   $g_{\bar{D}_s^*\Xi_c({}^2S_J)}$   & $-0.47-i0.00$ & $-13.90-i0.06$ & $-2.06-i1.87$ & $-$ & $-$ \\
   $g_{\bar{D}_s^*\Xi_c({}^2D_J)}$   & $-$ & $-$ & $-$                                & $0.04+i0.00$ & $-0.00-i0.04$ \\
   $g_{\bar{D}_s^*\Xi_c({}^4S_J)}$   & $-$ & $-$ & $-$                                & $-14.19-i0.01$ & $-$ \\
   $g_{\bar{D}_s^*\Xi_c({}^4D_J)}$   & $-0.01-i0.00$ & $0.12+i0.00$ & $-0.21-i0.18$   & $-0.04-i0.00$ & $0.08+i0.01$ \\
   $g_{\bar{D}_s\Xi_c^*({}^4S_J)}$   & $-$ & $-$ & $-$                                & $-10.54+i0.00$ & $-$ \\
   $g_{\bar{D}_s\Xi_c^*({}^4D_J)}$   & $-0.00-i0.00$ & $0.05+i0.00$ & $1.18+i0.30$    & $-0.05-i0.00$ & $-0.28-i0.19$ \\
   $g_{\bar{D}\Omega_c^*({}^4S_J)}$  & $-$ & $-$ & $-$                                & $14.69+i0.00$ & $-$ \\
   $g_{\bar{D}\Omega_c^*({}^4D_J)}$  & $0.00+i0.00$ & $-0.08-i0.00$ & $-0.09+i0.43$   & $0.08+i0.00$ & $1.52+i0.32$ \\
   $g_{\bar{D}_s^*\Xi_c'({}^2S_J)}$  & $-2.41-i0.01$ & $-6.32-i0.52$ & $3.74+i0.89$   & $-$ & $-$ \\
   $g_{\bar{D}_s^*\Xi_c'({}^2D_J)}$  & $-$ & $-$ & $-$                                & $-0.05+i0.00$ & $0.03-i0.04$ \\
   $g_{\bar{D}_s^*\Xi_c'({}^4S_J)}$  & $-$ & $-$ & $-$                                & $4.50-i0.00$ & $-$ \\
   $g_{\bar{D}_s^*\Xi_c'({}^4D_J)}$  & $-0.07-i0.00$ & $0.13+i0.01$ & $0.43-i0.31$    & $0.05-i0.00$ & $-0.11-i0.02$ \\
   $g_{\bar{D}^*\Omega_c({}^2S_J)}$  & $3.48+i0.01$ & $9.07+i0.74$ & $10.48+i3.81$    & $-$ & $-$ \\
   $g_{\bar{D}^*\Omega_c({}^2D_J)}$  & $-$ & $-$ & $-$                                & $0.09-i0.00$ & $-0.33-i0.01$ \\
   $g_{\bar{D}^*\Omega_c({}^4S_J)}$  & $-$ & $-$ & $-$                                & $-6.44+i0.00$ & $-$ \\
   $g_{\bar{D}^*\Omega_c({}^4D_J)}$  & $0.11+i0.00$ & $-0.22-i0.02$ & $0.00-i0.01$    & $-0.09+i0.00$ & $0.46+i0.05$ \\
   $g_{\bar{D}_s^*\Xi_c^*({}^2S_J)}$ & $-2.80-i0.01$ & $4.75-i0.63$ & $8.87-i0.52$    & $-$ & $-$ \\
   $g_{\bar{D}_s^*\Xi_c^*({}^2D_J)}$ & $-$ & $-$ & $-$                                & $0.09+i0.00$ & $0.00+i0.00$ \\
   $g_{\bar{D}_s^*\Xi_c^*({}^4S_J)}$ & $-$ & $-$ & $-$                                & $8.29+i0.00$ & $-$ \\
   $g_{\bar{D}_s^*\Xi_c^*({}^4D_J)}$ & $0.06+i0.00$ & $0.16-i0.01$ & $0.01-i0.01$     & $-0.13-i0.00$ & $-0.00-i0.00$ \\
   $g_{\bar{D}_s^*\Xi_c^*({}^6S_J)}$ & $-$ & $-$ & $-$                                & $-$ & $7.58+i2.43$ \\
   $g_{\bar{D}_s^*\Xi_c^*({}^6D_J)}$ & $0.16+i0.00$ & $0.07-i0.03$ & $0.13+i0.03$     & $0.04+i0.00$ & $0.00+i0.00$ \\
   $g_{\bar{D}^*\Omega_c^*({}^2S_J)}$& $4.12+i0.01$ & $-6.96+i0.91$ & $0.95-i1.30$    & $-$ & $-$ \\
   $g_{\bar{D}^*\Omega_c^*({}^2D_J)}$& $-$ & $-$ & $-$                                & $-0.14-i0.00$ & $-0.01-i0.00$ \\
   $g_{\bar{D}^*\Omega_c^*({}^4S_J)}$& $-$ & $-$ & $-$                                & $-12.17-i0.00$ & $-$ \\
   $g_{\bar{D}^*\Omega_c^*({}^4D_J)}$& $-0.10-i0.00$ & $-0.25+i0.02$ & $-0.03-i0.03$  & $0.22+i0.00$ & $0.01+i0.00$ \\
   $g_{\bar{D}^*\Omega_c^*({}^6S_J)}$& $-$ & $-$ & $-$                                & $-$ & $-8.02-i1.88$ \\
   $g_{\bar{D}^*\Omega_c^*({}^6D_J)}$& $-0.26-i0.00$ & $-0.13+i0.05$ & $0.08+i0.03$   & $-0.09-i0.00$ & $-0.02-i0.01$ \\
         \bottomrule
     \end{tabular*}
\end{table}

The first resonance is located almost at the
$\bar{D}_s\Xi_c$ threshold. Its pole position is determined to be
$M_{P_{c\bar{c}ss}} -i\Gamma/2=(4437.2-i0.002)$ MeV. Interestingly,
the width is tiny, which indicates that the $P_{c\bar{c}ss}(4437)$ is
almost stable. Since the $J/\psi \Xi$ threshold lies lower than this
resonance, it decays into $J/\psi$ and $\Xi$. Table~\ref{tab:2}
shows that the most dominant contribution to the first resonance come
from the $\bar{D}_s\Xi_c({}^{2}S_{1/2})$ channel. The next most
dominant one is the $\bar{D}^*\Omega_c^*({}^{2}S_{1/2})$ channel. The
$\bar{D}^*\Omega_c({}^{2}S_{1/2})$, $\bar{D}_s^*\Xi_c'
({}^{2}S_{1/2})$, and  $\bar{D}_s^*\Xi_c^*({}^{2}S_{1/2})$ channels
also come into significant play.  Note that the $D$-wave contributions
are small.  It is interesting to see that the 
$\bar{D}^*\Omega_c^*({}^{2}S_{1/2})$ channel is large, even though  
the corresponding threshold energy ($E_{\mathrm{th}}=4774$ MeV) is
quite higher than the mass of the $P_{c\bar{c}ss}(4437)$. Actually,
the transition amplitude for the $\bar{D}_s \Xi_c \to
\bar{D}^*\Omega_c^*$ process is larger than that for the $\bar{D}_s
\Xi_c \to \bar{D}\Omega_c$ transition. The reason is that while both
$K$- and $K^*$-exchange are involved in the $\bar{D}_s \Xi_c \to
\bar{D}^*\Omega_c^*$ transition, only $K^*$-exchange enters in the
$\bar{D}_s \Xi_c \to \bar{D}\Omega_c$. It also implies that 
$K$-exchange is more significant than $K^*$-exchange. 
The difference between the  $\bar{D}^*\Omega_c^*({}^{2}S_{1/2})$ and
 $\bar{D}_s^*\Xi_c^*({}^{2}S_{1/2})$ channels is related to the
 different values of the IS factor, as shown in Table~\ref{tab:1}.
Having examined the coupling strengths of the $P_{c\bar{c}ss}(4437)$
resonance to each channel, we draw a conclusion that it is not a
specific meson-baryon molecular state, because the four different
channels mentioned above are of the similar strengths. This is a
distinguished feature from the cases of the $P_{c\bar{c}}$ and
$P_{c\bar{c}s}$ resonances discussed in Refs.~\cite{Clymton:2024fbf,
  Clymton:2025hez}. 
 
The pole position of the second resonance with $J^P=1/2^-$ is obtained
to be $M_{P_{c\bar{c}ss}} -i\Gamma/2=(4504.1-i0.2)$ MeV, which lies
between the $\bar{D}_s\Xi_c$ and $\bar{D}_s\Xi_c'$ thresholds.
Table~\ref{tab:2} shows three channels most dominantly contribute to
this resonance in the following order: $\bar{D}\Omega_c$,
$\bar{D}_s^*\Xi_c$, and $\bar{D}_s\Xi_c'$ in the ${}^{2}S_{1/2}$ wave.
Another four channels in the ${}^{2}S_{1/2}$ wave are also minor
effects on the $P_{c\bar{c}ss}(4504)$: $\bar{D}^*\Omega_c$,
$\bar{D}^*\Omega_c^*$, $\bar{D}_s^*\Xi_c'$, and $\bar{D}_s^*\Xi_c^*$  
in the order of the magnitudes of the coupling strengths. Since the
$P_{c\bar{c}ss}(4504)$ is located below the $\bar{D}_s\Xi_c'$
threshold, we expect that the $\bar{D}_s\Xi_c'$ governs the formation
of the $P_{c\bar{c}ss}(4504)$. Interestingly, however, the
$\bar{D}\Omega_c$ channel is the most dominant one. Had we turned
off the contribution of this channel, the $P_{c\bar{c}ss}(4504)$
resonance would have disappeared, leaving only the 
enhancement in the $\bar{D}_s\Xi_c'$ threshold.
On the other hand, even though we remove the  $\bar{D}_s\Xi_c'$
channel, the $P_{c\bar{c}ss}(4504)$ survives. It indicates that it is
essential to consider the $\bar{D}\Omega_c$ channel for the
$P_{c\bar{c}ss}(4504)$. We have observed a similar situation in the
$h_1(1385)$ meson resonance~\cite{Clymton:2024pql}. Finally, we want
to mention that the $D$-wave contribution is tiny.  

The pole position of the third resonance with $J^P=1/2^-$ is given by  
$M_{P_{c\bar{c}ss}} -i\Gamma/2=(4703.7-i10.6)$ MeV in the complex
plane. Its width is the largest among the $P_{c\bar{c}ss}$'s with
negative parity. It is positioned below the $\bar{D}^*\Omega_c$
threshold. The $\bar{D}^*\Omega_c$ channel naturally provides the
largest contribution to form the $P_{c\bar{c}ss}(4704)$ resonance. The
$\bar{D}_s^*\Xi_c^*$ gives the next largest contribution. As observed
in Table~\ref{tab:2}, the channels below the $\bar{D}^*\Omega_c$
threshold, i.e., the $\bar{D}_s^*\Xi_c'({}^{2}S_{1/2})$,
$\bar{D}_s\Xi_c'({}^{2}S_{1/2})$, $\bar{D}_s^*\Xi_c({}^{2}S_{1/2})$,
and $\bar{D}_s\Xi_c^*({}^{4}D_{1/2})$ have non-negligible values of the
coupling strengths. This means that the $P_{c\bar{c}ss}(4704)$ can
decay into the corresponding states, which is the reason for the
relatively large width of the $P_{c\bar{c}ss}(4704)$.    

The upper-left panel of Fig.~\ref{fig:4} shows the
$P_{c\bar{c}ss}(4541)$ resonance with the pole position
$M_{P_{c\bar{c}ss}} -i\Gamma/2=(4541.3-i0.04)$ MeV. It lies below the
$\bar{D}_s \Xi_c'$ threshold. It is of great interest to see that
while the $\bar{D}_s \Xi_c'({}^2D_{3/2})$ channel does not contribute
to the $P_{c\bar{c}ss}(4541)$ resonance, seven different channels in
${}^4S_{3/2}$ wave above the $\bar{D}_s \Xi_c'$ threshold contribute
to it. The most dominant one comes from the $\bar{D}\Omega_c^*$
channel, and the next one, which is almost the same magnitude but has
the different sign, arises from the $\bar{D}_s^*\Xi_c$. Its small
width can be understood as follows: the $P_{c\bar{c}ss}(4541)$ cannot
decay into the $S$-wave because of the conservation of the total
angular momentum, it can only decay into meson-baryon states in
$D$-wave below the $\bar{D}_s \Xi_c'$ threshold. This orbital
excitation is physically not favorable, as shown in
Table~\ref{tab:2}. Note that it can also decay into the $J/\psi \Xi$
state in ${}^4S_{3/2}$-wave, though the coupling strength is small.    
Finally, we want to mention that there is a cusp structure just below
the $\bar{D}_s^*\Xi_c'$ threshold.

In the lower panel of Fig.~\ref{fig:4}, we draw the partial-wave
cross section with $J^P=5/2^-$, which illustrates a single resonance
below the $\bar{D}_s^*\Xi_c^*$ threshold. Its pole position is given
by $M_{P_{c\bar{c}ss}} -i\Gamma/2=(4756.5-i1.7)$ MeV. The
$\bar{D}^*\Omega_c^*({}^6S_{5/2})$ channel yields the most dominant
contribution. The next dominant one comes from the
$\bar{D}_s^*\Xi_c^*({}^6S_{5/2})$ one. Again, the small width of the
$P_{c\bar{c}ss} (4757)$ can be understood from the fact that it cannot
decay into lower $S$-wave meson-baryon states.

\begin{table}[htbp]
  \caption{\label{tab:3} Comparison of the results for the masses and
    widths of $P_{c\bar{c}ss}$'s in units of MeV.}
   \renewcommand{\arraystretch}{1.2}
  \begin{ruledtabular}
  \centering\begin{tabular}{ccccccccr}
   \multicolumn{3}{c}{Present work}  &
  \multicolumn{3}{c}{Ref.~\cite{Marse-Valera:2022khy}}
    & \multicolumn{3}{c}{Ref.~\cite{Roca:2024nsi}} \\ 
   $J^P$ & $M$ & $\Gamma$ & $J^P$ & $M$ & $\Gamma$ & $J^P$ & $M$ & $\Gamma$
   \\\hline
    $\frac{1}{2}^{-}$ & $4437.2$ & $0.004$&  &  &  &  &  &  \\
    $\frac{1}{2}^{-}$ & $4504.1$ & $0.4$   & $\frac{1}{2}^{-}$ &
    $4493.35$ & $73.67$ & $\frac{1}{2}^{-}$  & $4535$ & $9$ \\
    $\frac{3}{2}^{-}$ & $4541.3$ & $0.08$  &  &  & & $\frac{3}{2}^{-}$
                                        & $4602$  & $0$ \\ 
    $\frac{1}{2}^{-}$ & $4703.7$ & $21.2$ &
    $\frac{1}{2}^{-},\frac{3}{2}^{-}$ & $4633.38$ & $79.58$ &
    $\frac{1}{2}^{-},\frac{3}{2}^{-}$ & $4675$ & $10$  \\ 
    $\frac{5}{2}^{-}$ & $4756.5$ & $3.4$  &  &  & &
    $\frac{1}{2}^{-},\frac{3}{2}^{-},\frac{5}{2}^{-}$ & $4743$ & $0$  \\ 
  \end{tabular}
    \end{ruledtabular}
  \end{table}
It is interesting to compare the present results with those from other
works. In the second and third columns of Table~\ref{tab:3}, we list
the masses and widths of the negative-parity $P_{c\bar{c}ss}$ states
obtained in the present work. The fifth and sixth columns show the
corresponding results from Ref.\cite{Roca:2024nsi}, and the eighth and
ninth columns present those from Ref.\cite{Marse-Valera:2022khy}. Note
that in Ref.\cite{Roca:2024nsi}, two different sets of results were
presented, corresponding to different values of the chosen momentum
cutoff; for comparison, we take those with $600$MeV. As mentioned
previously, we predict five negative-parity $P_{c\bar{c}ss}$ states,
whereas Refs.\cite{Marse-Valera:2022khy, Roca:2024nsi} predict two and
four such states, respectively. In Ref.\cite{Roca:2024nsi}, the
$\bar{D}_s\Xi_c$ channel was neglected, which turns out to be important
for generating the first resonance in the present work. The existence
of $P_{c\bar{c}ss}(4437)$ would underscore the importance of the
$\bar{D}_s\Xi_c$ channel.

\begin{figure}[htbp]
   \centering
   \includegraphics[scale=0.9]{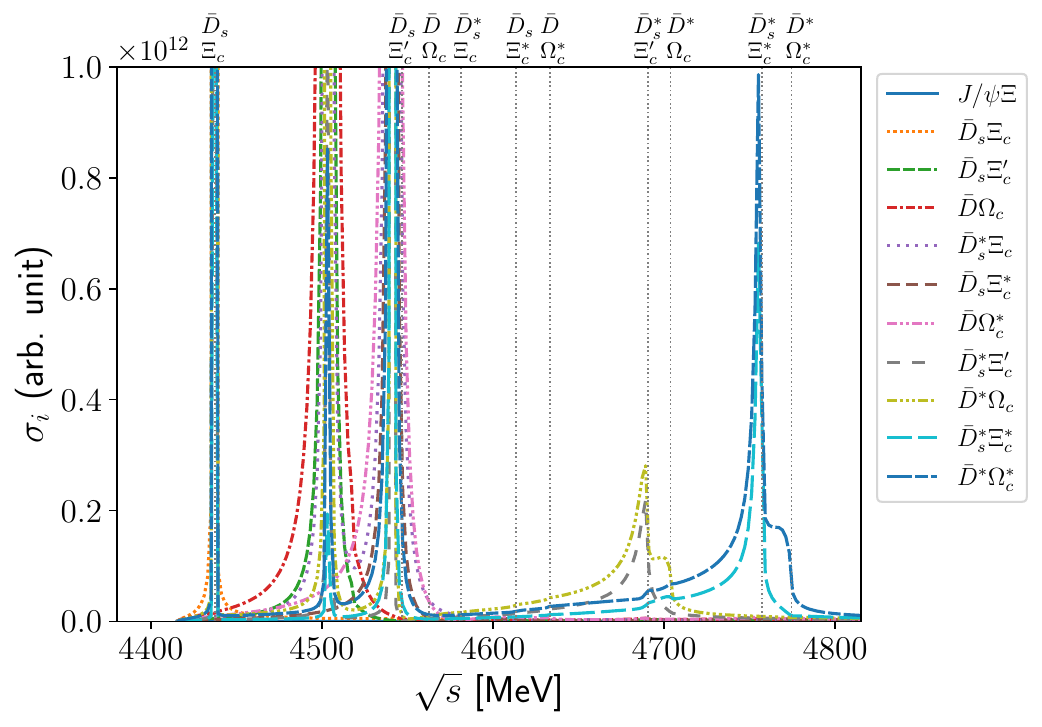}
   \caption{Results for the total cross sections for elastic
     scattering as functions of the CM energy.}    
   \label{fig:5}
\end{figure}
Figure~\ref{fig:5} shows the results for the total cross sections of
meson-baryon elastic scattering. While we seemingly observe the five
pentaquark resonances discussed above, it depends on which scattering
we consider. For example, we can see $P_{c\bar{c}ss}(4757)$ resonance
with $J^P=5/2^-$ only in the $\bar{D}_s^*\Xi_c^*$ and
$\bar{D}^*\Omega_c^*$ scattering processes. As for the
$P_{c\bar{c}ss}(4704)$ with $1/2^-$, we find it only in
$\bar{D}_s^*\Xi_c'$ and $\bar{D}^*\Omega_c$ processes. The reason can
be easily found by scrutinizing the coupling strengths listed in
Table~\ref{tab:2}. Note that we do not see any resonance signal in
$J/\psi \Xi$ scattering. We find similar features in the $J/\psi N$
and $J/\psi \Lambda$ interactions~\cite{Clymton:2024fbf,
  Clymton:2025hez}.

\subsection{Positive parity}
\begin{figure}[htbp]
   \centering
   \includegraphics[scale=0.5]{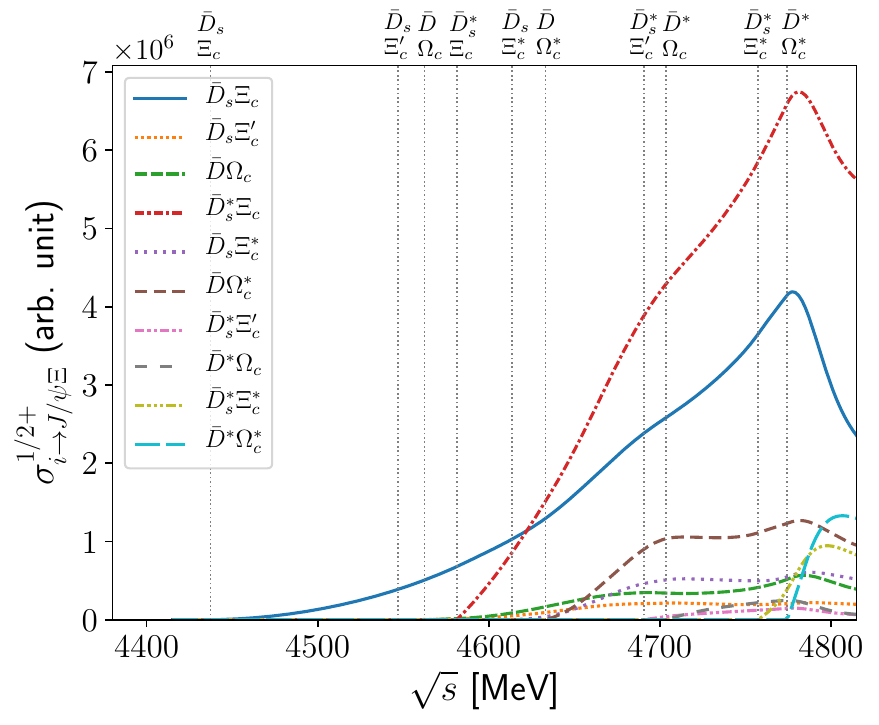}
   \includegraphics[scale=0.5]{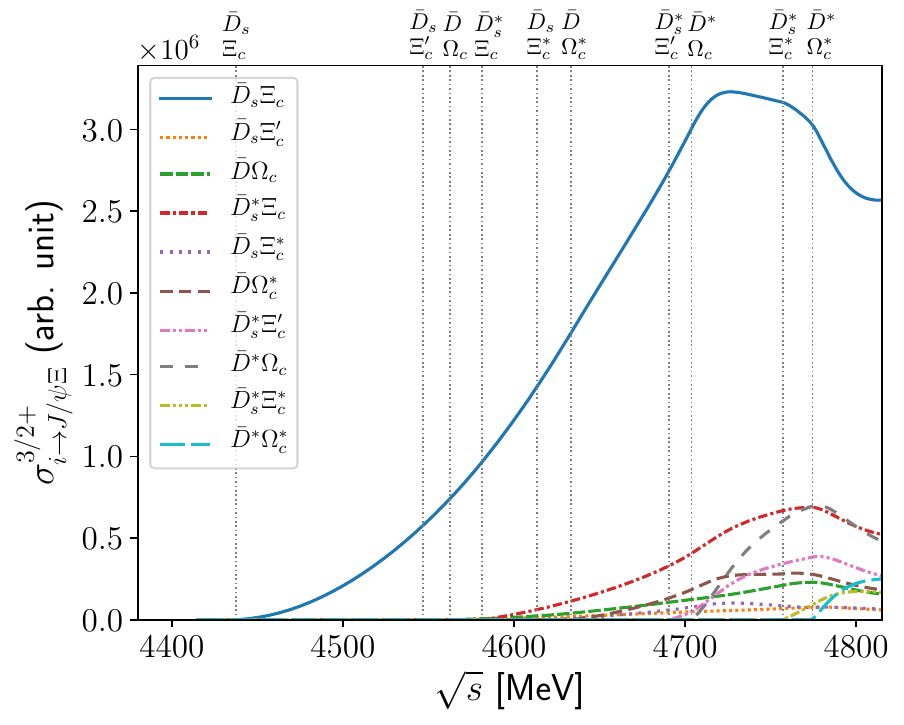}
   \includegraphics[scale=0.5]{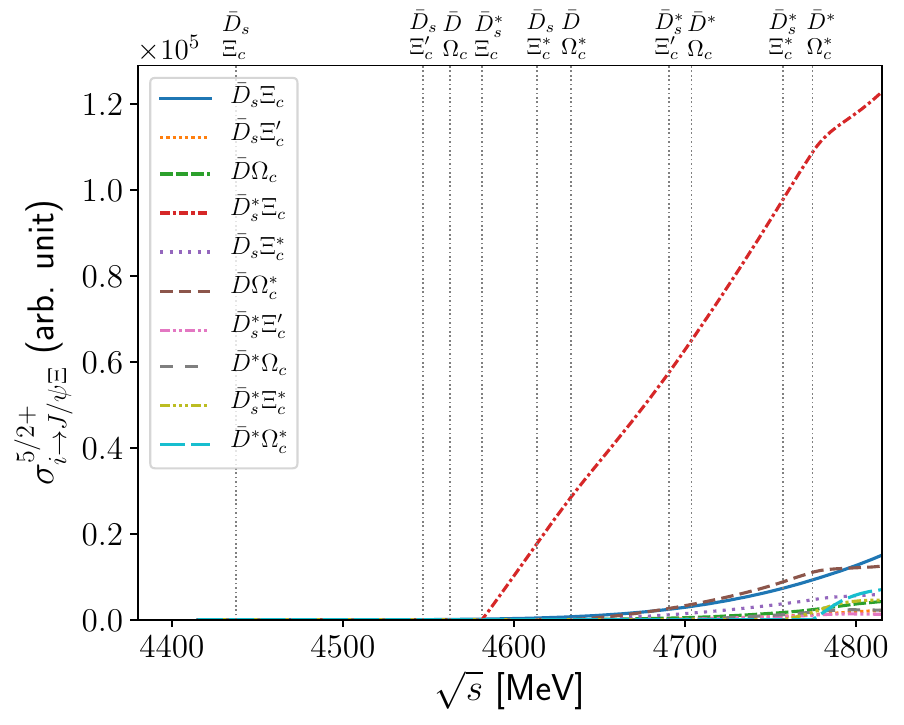}
   \caption{Center-of-mass energy dependence of partial-wave total
     cross sections for positive-parity states $(J=1/2, 3/2, 5/2)$
     corresponding to the spin-parity quantum numbers of
     $P_{c\bar{c}ss}$.}   
   \label{fig:6}
\end{figure}
\begin{table}[htbp]
 \caption{\label{tab:4}Coupling strengths of $P_{c\bar{c}ss}$'s with $J^P=1/2^+$ and $3/2^+$.}
 \centering\begin{tabular*}{\linewidth}{@{\extracolsep{\fill}} lrrr}
   \toprule
   $J^P$ & \multicolumn{2}{c}{$1/2^+$} & \multicolumn{1}{c}{$3/2^+$} \\
   $\sqrt{s_R}$[MeV] & $4665.6-i57.8$ & $4781.1-i25.7$ & $4712.3-i30.8$ \\
   \midrule
   $g_{J/\psi\Xi({}^2P_J)}$          & $0.03-i0.03$ & $0.16-i0.03$ & $0.02-i0.03$ \\
   $g_{J/\psi\Xi({}^4P_J)}$          & $0.24-i0.28$ & $0.28-i0.07$ & $0.05-i0.04$ \\
   $g_{J/\psi\Xi({}^4F_J)}$          & $-$ & $-$ & $0.00-i0.00$ \\
   $g_{\bar{D}_s\Xi_c({}^2P_J)}$     & $0.06-i1.06$ & $-1.41+i0.45$ & $1.06-i0.84$ \\
   $g_{\bar{D}_s\Xi_c'({}^2P_J)}$    & $2.01-i0.45$ & $0.55+i0.45$ & $-0.07+i0.28$ \\
   $g_{\bar{D}\Omega_c({}^2P_J)}$    & $-2.85+i0.56$ & $-1.02-i0.64$ & $0.09-i0.30$ \\
   $g_{\bar{D}_s^*\Xi_c({}^2P_J)}$   & $-0.23+i2.73$ & $-1.31+i0.95$ & $0.40-i0.76$ \\
   $g_{\bar{D}_s^*\Xi_c({}^4P_J)}$   & $-0.05+i3.41$ & $-0.78+i0.94$ & $-0.25+i1.45$ \\
   $g_{\bar{D}_s^*\Xi_c({}^4F_J)}$   & $-$ & $-$ & $-0.14-i0.01$ \\
   $g_{\bar{D}_s\Xi_c^*({}^4P_J)}$   & $-3.00+i4.07$ & $-0.56+i0.10$ & $-1.25+i0.27$ \\
   $g_{\bar{D}_s\Xi_c^*({}^4F_J)}$   & $-$ & $-$ & $0.22-i0.34$ \\
   $g_{\bar{D}\Omega_c^*({}^4P_J)}$  & $3.15-i5.87$ & $0.85+i0.11$ & $2.04-i0.44$ \\
   $g_{\bar{D}\Omega_c^*({}^4F_J)}$  & $-$ & $-$ & $-0.29+i0.26$ \\
   $g_{\bar{D}_s^*\Xi_c'({}^2P_J)}$  & $0.00+i0.00$ & $-0.96+i0.86$ & $-0.54+i2.70$ \\
   $g_{\bar{D}_s^*\Xi_c'({}^4P_J)}$  & $0.00+i0.00$ & $-0.19+i0.23$ & $2.65-i3.97$ \\
   $g_{\bar{D}_s^*\Xi_c'({}^4F_J)}$  & $-$ & $-$ & $-0.04+i0.07$ \\
   $g_{\bar{D}^*\Omega_c({}^2P_J)}$  & $0.00+i0.00$ & $1.37-i1.00$ & $0.48-i3.39$ \\
   $g_{\bar{D}^*\Omega_c({}^4P_J)}$  & $0.00+i0.00$ & $0.29-i0.48$ & $-2.94+i5.09$ \\
   $g_{\bar{D}^*\Omega_c({}^4F_J)}$  & $-$ & $-$ & $0.01-i0.06$ \\
   $g_{\bar{D}_s^*\Xi_c^*({}^2P_J)}$ & $0.00+i0.00$ & $1.95-i3.33$ & $0.00+i0.00$ \\
   $g_{\bar{D}_s^*\Xi_c^*({}^4P_J)}$ & $0.00+i0.00$ & $1.43-i1.79$ & $0.00+i0.00$ \\
   $g_{\bar{D}_s^*\Xi_c^*({}^4F_J)}$ & $-$ & $-$ & $0.00+i0.00$ \\
   $g_{\bar{D}_s^*\Xi_c^*({}^6P_J)}$ & $-$ & $-$ & $0.00+i0.00$ \\
   $g_{\bar{D}_s^*\Xi_c^*({}^6F_J)}$ & $0.00+i0.00$ & $0.01-i0.09$ & $0.00+i0.00$ \\
   $g_{\bar{D}^*\Omega_c^*({}^2P_J)}$& $0.00+i0.00$ & $-2.00+4.24i$ & $0.00+i0.00$ \\
   $g_{\bar{D}^*\Omega_c^*({}^4P_J)}$& $0.00+i0.00$ & $-1.79+i2.64$ & $0.00+i0.00$ \\
   $g_{\bar{D}^*\Omega_c^*({}^4F_J)}$& $-$ & $-$ & $0.00+i0.00$ \\
   $g_{\bar{D}^*\Omega_c^*({}^6P_J)}$& $-$ & $-$ & $0.00+i0.00$ \\
   $g_{\bar{D}^*\Omega_c^*({}^6F_J)}$& $0.00+i0.00$ & $0.02+i0.05$ & $0.00+i0.00$ \\
  \bottomrule
 \end{tabular*}
\end{table}
A significant feature of the current off-shell coupled channel
formalism is that it also predicts the positive-parity pentaquark
states coming from the $P$-wave contributions, as already examined in
the the production mechanism of the $P_{c\bar{c}}$ and $P_{c\bar{c}s}$
pentaquark states~\cite{Clymton:2024fbf, Clymton:2025hez}. Similarly,
we also observe three double-strangeness hidden-charm resonances with
positive parity. Figure~\ref{fig:5} depicts the results for the $i\to
J/\psi \Xi$ transition partial-wave cross sections with positive
parity. In Table~\ref{tab:4}, we list the results for the coupling
strengths of the positive-parity $P_{c\bar{c}ss}$ pentaquark
resonances to the channels involved. In the upper-left panel of Fig.~\ref{fig:5}, we
observe a broad resonant structure, of which the pole position is at
$(4665.6 - i57.8)$ MeV. It is located between the $\bar{D}\Omega_c^*$
and $\bar{D}_s^*\Xi_c'$ thresholds. In Table~\ref{tab:4}, we find that
four different channels, i.e., $\bar{D}\Omega_c^*({}^4 P_{1/2})$,
$\bar{D}_s\Xi_c^*({}^4 P_{1/2})$, $\bar{D}\Omega_c({}^2 P_{1/2})$, and
$\bar{D}_s\Xi_c'({}^2 P_{1/2})$ channels mainly contribute to form the
$P_{c\bar{c}ss}(4666)$ resonance with $J^P=1/2^+$. The pole position
of the second resonance $P_{c\bar{c}ss}(4781)$ is obtained to be
$(4781.1-i25.7)$ MeV. Interestingly, many different channels
contribute to it, as shown in Table~\ref{tab:4}. The third resonance
$P_{c\bar{c}ss}(4712)$ with $J^P=3/2^+$ can be seen in the
upper-right panel of Fig.~\ref{fig:6}, which is located at
$(4712.3-i30.8)$ MeV. The three different channels primarily govern
the production of the $P_{c\bar{c}ss}(4712)$. There is no pentaquark
resonance with $J^P=5/2^+$ as depicted in the lower panel of
Fig.~\ref{fig:6}, since only the $\bar{D}_s^*\Xi_c$ channel is large,
whereas all other channels are suppressed.  
\subsection{Uncertainties of the cutoff masses}
\begin{table}[htbp]
    \caption{\label{tab:5} The change in pole position due to the
      change of $\Lambda_0=\Lambda-m$. The pole position in MeV unit.} 
   \begin{ruledtabular}
    \centering
    \begin{tabular}{lrrr}
     \multirow{2}{*}{$J^P$} &  
     \multicolumn{3}{c}{$\Delta\Lambda_0/\Lambda_0$} \\
     & $0\%$ & $-10\%$ & $+10\%$
     \\\hline
       $1/2^{-}$ 
       & $4437.2-i0.002$ & cusp & $4420.8-i0.01$ \\
       & $4504.1-i0.2$ & $4531.5-i0.1$ & $4466.0-i0.3$ \\
       & $4703.7-i10.6$ & cusp & $4633.2-i4.0$ \\
       $3/2^{-}$ 
       & $4541.3-i0.04$ & $4570.8-i0.02$ & $4323.77-i0.04$ \\
       $5/2^{-}$ 
       & $4756.5-i1.7$ & cusp & $4750.1-i3.2$ \\
       $1/2^{+}$ 
       & $4665.6-i57.8$ & $4665.2-i70.6$ & $4662.8-i46.0$ \\
       & $4781.1-i25.7$ & $4782.5-i33.0$ & $4777.9-i19.5$ \\
       $3/2^{+}$ 
       & $4712.3-i30.8$ & $4708.6-i38.9$ & $4711.9-i22.4$ \\
      \end{tabular}
   \end{ruledtabular}
 \end{table}
Since there are no experimental data on the double-strangeness
hidden-charm pentaquarks, we have no experimental guideline to fix the
cutoff masses. Thus, we choose the reduced cutoff mass to be
$\Lambda_0 = \Lambda-m=700$ MeV, as in the previous
studies~\cite{Clymton:2024fbf, Clymton:2025hez}. We examine the
uncertainties that are caused by this value. Table~\ref{tab:5}
demonstrates how the pole positions of the pentaquark resonances are
shifted as $\Lambda_0$ is varied by 10\%. When $\Lambda_0$ is
increased, the masses of the pentaquark states with negative parity
decrease whereas their widths are enlarged. On the other hand, when
$\Lambda_0$ is decreased by 10\%, the results are changed to the
opposite direction. Moreover, $P_{c\bar{c}ss}(4437)$,
$P_{c\bar{c}ss}(4704)$ and $P_{c\bar{c}ss}(4757)$ resonances are
reduced to the cusp structures. The reason lies in the fact that by
employing values of the cutoff masses softer than 700 MeV, the overall
transition amplitudes are more suppressed. 
However, the masses of those with positive parity are slightly
changed as $\Lambda_0$ is varied, whereas their widths are altered by
about 20-30\%. Future experimental data will clarify these
uncertainties.  

\section{Summary and conclusion\label{sec:4}}
In this work, we have investigated the possible existence of
double-strangeness hidden-charm pentaquark states, denoted as
$P_{c\bar{c}ss}$, using an off-shell coupled-channel formalism. We
constructed eleven meson--baryon channels with total strangeness $S =
-2$, combining charmed mesons and singly charmed baryons. The
interaction kernel was derived from an effective Lagrangian that
respects heavy-quark spin symmetry, hidden local symmetry, and flavor
SU(3) symmetry. The coupled-channel scattering equations were solved
in a partial-wave basis using the Blankenbecler--Sugar reduction
scheme. 
We found five $P_{c\bar{c}ss}$ resonances with negative parity. Among
them, three states have spin-parity $J^P = 1/2^-$, one has $J^P =
3/2^-$, and one has $J^P = 5/2^-$. All of these negative-parity states
are located below their respective thresholds and are dynamically
generated. The dominant channels contributing to each resonance were
identified through the analysis of their coupling strengths. The
narrowest state is found near the $\bar{D}_s \Xi_c$ threshold, while
broader states appear closer to higher thresholds such as
$\bar{D}^*\Omega_c$. 
In addition, three positive-parity $P_{c\bar{c}ss}$ states are found:
two with spin-parity $J^P = 1/2^+$ and one with $J^P = 3/2^+$. These
states are located above the corresponding thresholds and exhibit
substantial widths. Their coupling patterns indicate significant
contributions from multiple meson--baryon channels, especially those
involving $P$-wave interactions. 

The theoretical uncertainties associated with the cutoff mass were
also examined by varying the reduced cutoff parameter $\Lambda_0 =
\Lambda - m$ by $\pm 10\%$. This variation resulted in modest shifts
in pole positions and in some cases transformed resonances into
cusp-like structures, but the overall spectrum and structure remained
qualitatively stable. 
The present study provides theoretical predictions for the mass
spectrum, widths, and dominant coupling channels of the
double-strangeness hidden-charm pentaquark states. These predictions
may serve as useful input for future experimental searches in
processes involving the $J/\psi \Xi$ final state. 
The triple-strangeness hidden-charm pentaquark states can also be
studied within the same framework, which is under investigation.

\begin{acknowledgments}
The present work was supported by the Young Scientist Training (YST)
Program at the Asia Pacific Center for Theoretical Physics (APCTP)
through the Science and Technology Promotion Fund and Lottery Fund of 
the Korean Government and also by the Korean Local Governments –
Gyeongsangbuk-do Province and Pohang City (SC), the Basic Science
Research Program through the National Research Foundation of Korea
funded by the Korean government (Ministry of Education, Science and
Technology, MEST), Grant-No. RS-2025-00513982 (HChK), and the PUTI Q1 
Grant from University of Indonesia under contract
No. NKB-441/UN2.RST/HKP.05.00/2024 (TM).
\end{acknowledgments}

\bibliography{Pcss}
\bibliographystyle{apsrev4-2}

\end{document}